\documentclass[11pt]{article}
\usepackage{macros}

\usepackage{amsthm,amsmath,amssymb,comment}
\usepackage{graphicx}
\usepackage{setspace}
\usepackage{amsfonts}
\usepackage{authblk}
\usepackage{natbib}
\usepackage{titletoc}
\usepackage{booktabs}
\usepackage{fullpage}
\usepackage{hyperref}
\usepackage{todonotes}
\usepackage{multirow}
\usepackage{tabularx}

\usepackage{pdflscape}
\theoremstyle{plain}

\newtheorem{remark}{Remark}[section]
\theoremstyle{definition}

\newtheorem{mc}{Monte Carlo design}
\crefname{mc}{Monte Carlo design}{Monte Carlo designs}
\usepackage{clipboard}
\newclipboard{draft}

\newcommand{\ismd}{\textsf{P-ISMD }}
\newcommand{\osmd}{\textsf{OP-OSMD }}
\newcommand{\is}{\textsf{IS }}
\newcommand{\es}{\textsf{ES }}

\newcommand{\ismdns}{\textsf{P-ISMD}} %
\newcommand{\osmdns}{\textsf{OP-OSMD}} %
\newcommand{\isns}{\textsf{IS}} %
\newcommand{\esns}{\textsf{ES}} %

\begin{document}

\begin{titlepage}
\footnotesize
    \title{Efficient Estimation of Average Derivatives in NPIV Models:\\ Simulation Comparisons of Neural Network Estimators\thanks{%
We are grateful to Vanya Klenovskiy for excellent research assistance in applying our code
to estimate the Strawberry demand and to Josh Purtell in partially checking our code
documentation. We thank Denis Chetverikov for insightful discussions at the Gary
Chamberlain online seminar. We also thank A. Babii, S. Bonhomme, E. Ghysels, S. Han, G.
Imbens, P. Kline, O. Linton, W. Newey, M. Pelger, D. Ritzwoller, B. Ross, P. Sant'Anna, Y.
Sun, A. Timmermann, D. Xiu, Y. Zhu, and participants at various seminars and conferences
for helpful comments. Any errors are the responsibility of the authors. An implementation
for the procedures is available at \url{https://github.com/jiafengkevinchen/cct-ann}.}}
\author{Jiafeng Chen\thanks{Harvard Business School and Department of Economics, Harvard
University. \textit{Email}:
jiafengchen@g.harvard.edu.}\qquad Xiaohong Chen\thanks{%
Cowles Foundation for Research in Economics, Yale
University. \textit{Email}: xiaohong.chen@yale.edu.}\qquad Elie Tamer\thanks{%
Department of Economics, Harvard University. \textit{Email}:
elietamer@fas.harvard.edu.}}
\date{First draft: September 2019 -  Revised draft: September 2022
}
\maketitle

\singlespacing
\begin{abstract}
\footnotesize Artificial Neural Networks (ANNs) can be viewed as \emph{nonlinear sieves}
that can approximate complex functions of high dimensional variables more effectively than
linear sieves. We investigate the performance of various ANNs in
nonparametric instrumental variables (NPIV) models of moderately high dimensional
covariates that are relevant to empirical economics. We present two efficient procedures
for estimation and inference on a weighted average derivative (WAD): an
\textbf{o}rthogonalized \textbf{p}lug-in with \textbf{o}ptimally-weighted \textbf{s}ieve
\textbf{m}inimum \textbf{d}istance (OP-OSMD) procedure and a sieve efficient score (ES)
procedure. Both estimators for WAD use ANN sieves to approximate the unknown NPIV function
and are root-$n$ asymptotically normal and first-order equivalent. We provide a detailed
practitioner's recipe for implementing both efficient procedures. 
We compare their
finite-sample performances in various simulation designs that involve smooth NPIV function
of up to 13 continuous covariates, different nonlinearities and covariate correlations.
Some Monte Carlo findings include: 1) tuning and optimization are more delicate in ANN
estimation; 2) given proper tuning, both ANN estimators with various architectures can
perform well; 3) easier to tune ANN OP-OSMD estimators than ANN ES estimators; 4) stable
inferences are more difficult to achieve with ANN (than spline) estimators; 5) there are
gaps between current implementations and approximation theories. Finally, we apply ANN
NPIV to estimate average partial derivatives in two empirical demand examples with
multivariate covariates.

\medskip \medskip

\noindent \emph{JEL Classification:} C14; C22

\medskip

\noindent \emph{Keywords:} Artificial neural networks; Relu; Sigmoid; Nonparametric instrumental variables; Weighted average derivatives; Optimal sieve minimum distance; Efficient influence; Semiparametric efficiency; Endogenous demand.
\end{abstract}

\end{titlepage}

\baselineskip=18pt

\section{Introduction}

Deep layer Artificial Neural Networks (ANNs) are increasingly popular in machine learning
(ML), statistics, business, finance, and other fields. The universal approximation
property of a variety of ANN architectures has been established by \cite*
{hornik1989multilayer} and many others. Early on, computational difficulties have hindered
the wide applicability of ANNs. Recently,  improvements in computing have led to
successful applications of deep layer ANNs in computer vision, natural language processing
and other areas, with complex nonlinear relations among many covariates and large data
sets of high quality.\footnote{By high quality we mean data sets with very high
signal-to-noise ratios. Unfortunately, many economic and social science data sets have low
signal-to-noise ratios.} Many problems where deep layer ANNs are extremely effective
involve prediction problems (i.e. estimating conditional means or densities)---or problems
in which nuisance parameters are themselves predictions. Recently, \cite{farrell2018deep}
and \cite*{athey2019using}, among others, have applied multi-layer ReLU ANNs to estimate
average treatment effects under unconfoundedness and demonstrated their good performance
in estimating unknown conditional means and densities of multivariate covariates.%
It remains to be seen whether ANNs are similarly effective for structural estimation
problems with nonparametric endogeneity.

To that end, we consider semiparametric efficient estimation and inference for a weighted
average (partial) derivative (WAD) of a nonparametric instrumental variables regression
(NPIV) via ANN sieves. Specifically, we assume an unknown structure function $h$ satisfies
the NPIV model: $\E[Y_{1} - h(Y_2) \mid X] = 0$, where $Y_2$ is a continuous random vector
of moderately high dimension (including endogenous regressors that are excluded from $X$),
and $X$ is a vector of moderately high dimensional conditioning variables. We are
interested in efficient estimation and inference for a WAD parameter of the smooth NPIV
function $h (Y_2)$, without sparsity assumptions on $h(\cdot)$.\footnote{Of course, in
lieu of sparsity, we do require smoothness assumptions.}%
WADs of structural relationships are linked to elasticities of endogenous demand systems
in economics. It is essentially a treatment effect parameter under confounding and
endogenous continuous treatment. Although there is a large literature on efficient
estimation of the average treatment effect and other causal parameters under
unconfoundedness, there are far fewer results on efficient estimation and inference on the
average treatment effect in nonparametric models with endogenous continuous treatment.

This paper makes three contributions. 
First, we present two classes of efficient estimators for WADs of NPIV models where
unknown $h_0(Y_2)$ is approximated by ANN sieves: the optimally weighted sieve minimum
distance estimators and the efficient score-based estimators.%
Under some regularity conditions both types of estimators are root-$n$ asymptotically
normal, semiparametrically efficient, and hence are first-order equivalent. Second, we
detail a {\it practitioner's recipe} that include a step by step guide for implementing
these two classes of estimators. Third, and perhaps most importantly, we present a large
set of Monte Carlo results on finite-sample performances of various ANN estimators. These
are implemented using increasingly complex designs, such as NPIV function containing up to
13 continuous covariates (including endogenous regressors), various nonlinearities and
correlations among the covariates.

We now briefly introduce the two classes of efficient estimation procedures that we consider. Both procedures are inspired by the semiparametric efficiency bound characterization in \cite{ai2012semiparametric} (henceforth AC12) for the WAD of the unknown $h(Y_2)$ in a NPIV model $\E[Y_{1}-h(Y_{2}) \mid X]=0$. The first procedure is based on minimizing an optimal criterion, the optimally-weighted orthogonalized \emph{sieve minimum distance} (SMD) criterion. 
This procedure is numerically equivalent to a semiparametric two-step procedure, where
the unknown NPIV function $h (\cdot)$ is
estimated via an optimally weighted SMD in the first step, and the WAD of $h(\cdot)$ is estimated using a
sample analogue of an orthogonalized unconditional moment \citep{chamberlain1992jbes} in the second step, with the unknown $h$ substituted by the optimally weighted SMD estimator from the first step. This  will be denoted
as \osmd in our paper. AC12 already introduced this procedure and presented a small Monte
Carlo study demonstrating its finite-sample performance using a spline SMD in the first
step when the unknown $h(\cdot)$ is a function of a scalar endogenous variable
$Y_2$. It is unclear how this procedure will perform when $Y_2$ could be a
continuous random vector of higher dimension and when $h(\cdot)$ is approximated via a
neural network.

The second procedure is based on the efficient score (equivalently, efficient influence
function).\footnote{The efficient score/influence function approach to efficient
estimation has a long history in semiparametrics. See, e.g., \cite{bickel1993efficient},
\cite{pfanzagl1982lecture},
Section 25.8 of \cite{van2000asymptotic} and references therein, for an introduction.}
AC12 derived a characterization of the efficient influence (or equivalently, efficient
score) for the WAD of a NPIV model. It is also the asymptotic influence function of the
\osmd estimator.\footnote{This is not surprising since the efficient influence function is
unique.} The efficient influence is the sum of the orthogonalized unconditional moment
(the one used for the  \osmd   estimator) and an adjustment term accounting for
plugging-in estimated $h(Y_2)$, often referred to as the Riesz representer term. Compared
to simpler settings, e.g. estimating average treatment effect under unconfoundedness, the
Riesz representer term here has no closed-form expression, but is characterized as one
solution to an optimization problem over an infinite-dimensional Hilbert space induced by
a norm connected to the optimally weighted minimum distance objective. The components of
the efficient influence function can nonetheless be consistently estimated via sieve
approximations. The procedure using the sample estimated efficient influence (i.e.,
efficient score) will be denoted as \es in our paper. To the best of our knowledge, there
is no published work on theory or simulation on the finite-sample performance of any \es
estimator for the WAD in a NPIV model yet.

In this paper we investigate the finite-sample performance of both efficient procedures
when the unknown function $h(Y_2)$ is estimated via various ANN SMDs and when $h (Y_2)$
depends on moderately high dimensional continuous regressors $Y_2$ (some of which are
endogenous). We describe some stylized findings from our simulations. Our simulations
reveal that the ANN \osmd is more stable and easier to
implement than ANN  \es for estimation of the average (partial) derivative in a NPIV model
with unknown conditional variance $\Sigma(X)\equiv \var(Y_1-h(Y_2) \mid X )$.%

In practice, it could be appealing to report simpler inefficient estimators that are still
consistent and $\sqrt{n}$-asymptotically normal. It is also possible that computationally
simpler inefficient estimators may perform better than the efficient estimators in finite
samples, as the efficient estimators often require estimating additional nuisance
parameters. For the sake of comparison, we include two first-order asymptotically
equivalent inefficient estimators of the WAD of a NPIV function, denoted by  \ismd and
\isns. The \ismd is a simple plug-in identity-weighted SMD estimator that was proposed in
\cite {ai2007estimation} (henceforth AC07). The \is is what we call ``inefficient score''
estimator that is based on sample analog of the asymptotic influence function of the \ismd
estimator (derived in AC07).\footnote{Different inefficient estimators of the WAD can have
different asymptotic influence functions and hence different asymptotic variances. That is
why we define the \is estimator based on the asymptotic influence function of the \ismd
estimator of AC07, so that they will have the same asymptotic variance.} We note that both
\ismd and \is are asymptotically efficient for a WAD of a nonparametric regression $\E[Y_1
\mid Y_2]$, in the absence of endogeneity. However, they are no longer efficient for the
WAD of a NPIV function $h (Y_2)$ identified by the conditional moment restriction
$\E[Y_{1}-h(Y_{2}) \mid X]= 0$ (for $Y_2\neq X$).

We compare the finite sample performance of these efficient (\osmdns, \esns) and
inefficient (\ismdns, \isns) estimation procedures in four Monte Carlo designs with
moderate sample sizes ($n=1000$ to $n=10000$).\footnote{Since both score-based estimators
\es and \is are based on orthogonal moments, we also provide comparison with their
cross-fitted versions. The cross-fitting orthogonal moments estimators have become very
popular following%
\citep*{chernozhukov2018double,chernozhukov2021locally} and others, although no published
work has applied cross-fit to efficient estimation of WAD in NPIV yet.} In
\cref{mc:simple}, we estimate a simple nonparametric regression and two-stage least
squares data-generating process, as a useful baseline. In \cref{mc:2,mc:3}, we estimate
the average partial derivative of a NPIV function $h(Y_2)$ with respect to an endogenous
variables using various ANN sieves and spline sieves. In \cref{mc:calibrated}, we
calibrate a data-generating process to the gasoline empirical application
\citep{blundell2012measuring}, and repeat the exercises for \cref{mc:2,mc:3}.

Our Monte Carlo experiments allow for comparisons along several dimensions:
\begin{itemize}
   \item For ANN estimators, how much does ANN architecture (activation, depth, width)
    matter? How much do other tuning parameters matter?
    \item Across types of estimation procedures, how do ANN SMD estimators compare to
    ANN score estimators, along with alternative procedures like adversarial GMM
    \citep{dikkala2020minimax}?
    \item Within a type of estimation procedure, do ANN estimators exhibit superior
    finite-sample performance compared to linear sieve (e.g., spline) estimators, when
    dimension of $Y_2$ is moderately high?\footnote{To be clear, we are not speaking of
    ``high dimension'' in the $\dim(Y_2)/n \not\to0$ sense.}
\end{itemize}
The main, stylized takeaways from our Monte Carlo experiments are as follows: 

\begin{itemize}
 \item Choices of hyperparameters in optimization---learning rate,
 stopping criterion---are delicate and can affect performance of ANN-based
 estimators. Nonconvex optimization could lead to unstable performances. 
 However, certain values of the hyperparameters do result in good performance of ANN based estimators.
 
\item We do not empirically observe systematic differences in finite-sample performances 
as a function of ANN architecture, within the feedforward neural network family. In our
experience,  ANN architecture is not as important as tuning the optimization procedure.

\item Stable inferences are currently more difficult to achieve for ANN based estimators
for models with nonparametric endogeneity.

\item ANN \osmd and ANN \is have smaller biases than ANN \ismd  for the average derivative
parameter.

\item ANN \es and ANN cross-fitted \es are sensitive (in terms of bias) to the estimation
of the optimal weighting $\Sigma^{-1}(X)$ in Riesz representer adjustment part. ANN \osmd
  
\item Spline \osmd, spline \ismd, spline \is and spline \es for the average derivative
  parameter are less biased, stable and accurate, and can outperform their ANN
  counterparts, even when the NPIV function $h(Y_2)$ depends on moderately
  high-dimensional continuous covariates $Y_2$ (as high as thirteen in the simulation
  studies).

  \item Generally, there seems to be gaps between intuitions suggested by
  approximation theory and current implementation.
\end{itemize}

Lastly, as applications to real data, we apply ANN sieve NPIV to estimate average price
elasticity of a gasoline demand using the data set of \cite*{blundell2012measuring}, and
to estimate average derivatives of a price-quantity relation in differentiated product
markets using the data set of \cite{compiani2019market}. Both applications involve
nonparametric structure functions of multi-dimensional covariates (including endogenous
price), and our ANN applications do not impose any semiparametric shape restrictions.

\paragraph{Related literature on ANN NPIVs.}
We view various ANNs as examples of \textit{nonlinear} sieves, which, compared to linear
sieves, can have faster approximation error rates for large classes of nonlinear functions
of high dimensional regressors. Once after the approximation error rate of a specific ANN
sieve is established for a class of unknown functions,\footnote{Different ANN sieves have
different approximation error rates for different function classes. See, for example,
\cite{barron1993universal} and \cite{chen1999improved} for approximation errors rates for
single hidden layer ANNs for Barron class; \cite {yarotsky2017error},
\cite{shen2021optimal}, \cite{shen2021neural} for approximation error rates of multi layer
ReLU ANNs for typical smooth function class; \cite{schmidt2019deep} for approximation
error rates of deep layer ReLU ANNs for composition function classes.} the asymptotic
properties of estimation and inference based on the ANN sieve could be established by
applying the general theory of sieve-based methods. The nonparametric convergence rates in
\cite {ai2003efficient,ai2007estimation} (henceforth, AC03, AC07) explicitly allow for
nonlinear sieves such as ANNs to approximate and estimate the unknown structure functions
of endogenous variables. They establish the root-$n$ asymptotic normality of regular
functionals of nonparametric conditional moment restrictions with smooth residual
functions. Due to the small sample size and computational limitation, earlier applications
in econometrics have focused on single-hidden layer ANNs. For instance,
\cite{chen2009land} applied single hidden layer sigmoid ANN SMD to estimate the unknown
habit function in a semi-nonparametric asset pricing conditional moment model with a time
series sample size of about 200 quarterly observations. To the best of our knowledge,
\cite*{hartford2017deep} is the first paper to apply multi-layer (2 hidden layer) ANNs to
estimate an NPIV structural function. Since then, numerous studies have followed up, see
\cite*{dikkala2020minimax} and the references there in.\footnote{However, as documented in
our simulation results, the WAD parameter estimated via plugging in the estimated $h
(\cdot)$ via adversarial GMM \citep{dikkala2020minimax} can be biased. This is not
surprising since the tuning parameter choice for nonparametric estimation of $h(\cdot)$
may be different from that for the efficient estimation of the WAD.}  In a project that
started after our first draft, \cite*{chen2020nn2} established rate of convergence for
multi-layer ANN optimally weighted SMD estimation of general nonparametric conditional
moment restrictions for time series data, and proposed ANN sieve quasi-likelihood ratio
inference for possibly slower-than-root-$n$ estimable linear functionals. However, they do
not consider efficient estimation for root-$n$ estimable linear functionals of NPIV such
as the WAD parameter, which is our parameter of interest.

Our simulation studies and empirical applications indicate that, although multi-layer ANNs
can perform well after careful choice of tuning parameters, they have no clear advantage
over single hidden layer ANNs or spline sieves for efficient estimation of WAD in a NPIV
model when the unknown structure function $h(Y_2)$ is a relatively smooth function of
multi-dimensional $Y_2$, which is likely the case in economic endogenous demand
estimation. Just like the simulation paper by \cite{lee1993testing} about the performance
of single-hidden layer ANNs on testing nonlinear regression models, our paper documents
that ANNs can also be one promising tool in efficient estimation and inference for causal

The rest of the paper is organized as follows. \Cref{sec:model} introduces the model, and
the two classes of efficient estimation procedures. \Cref{sec:implementation} provides
implementation details for all the estimators considered in the Monte Carlo studies.
\Cref{sec:MC} contains three simulation studies and detailed Monte Carlo comparisons of
various ANN and spline based estimators. \Cref{sec:application} presents two empirical
illustrations and \Cref {sec:conclusion} concludes.

\section{Efficient Estimation Procedures for Average Derivatives in NPIV Models}
\label{sec:model}

We first present the model and recall the semiparametric efficiency bound characterization. We then present two classes of efficient estimation procedures.

We are interested in semiparametrically efficient estimation of the average
partial derivative:
\begin{equation*}
\theta_0 \equiv \E[a(Y_2) \nabla_1 h_0 (Y_2 ))],
\end{equation*}
 where $a(\cdot)$ is a known positive weight function, $\nabla_1$ is
 the partial derivative w.r.t. the first argument and the unknown real-valued function
 $h_0 \in \mathcal{H}$ is identified via a conditional moment
 restriction\footnote{See, e.g.,  \cite{NP2003instrumental}, \cite{BCK2007}, \cite{andrews2017examples}
 for identification of a NPIV model.}
\begin{equation}
\E[Y_1 - h_0(Y_2) \mid X] = 0,\quad 
\text{$X$ almost sure} \label{ac:seq}
\end{equation}
Previously, 
\cite{ai2007estimation} (AC07)  presented a root-$n$ consistent asymptotically normally distributed identity-weighted SMD estimator of $\theta_0$, nonlinear sieves such as single hidden layer ANN sieve is allowed for in their sufficient conditions.
AC12 presented the semiparametric efficiency bound of $\theta_0$ and an efficient estimator based on orthogonalized optimally weighted SMD (see their section 4.2).\footnote{\cite{ai2012semiparametric} derived the efficiency bound via the
\textquotedblleft orthogonalized residual\textquotedblright\ approach, which extends the earlier work of \cite{chamberlain1992jbes} to allow for unknown functions entering a system of sequential moment restrictions.} 
\cite{ECO-019} presented
efficiency bound calculation for average weighted derivatives of a NPIV
model without assuming point identification of the NPIV function, but
pointed out that the $\sqrt{n}$-asymptotically normal estimator of linear
functionals of NPIV in \cite{santos} fails to achieve the efficiency bound.
\cite*{chenpouzopowell} proposed efficient estimation of
weighted average derivatives of nonparametric quantile IV regression via
penalized linear sieve GEL procedure, without providing any simulation results on how their procedure performs in finite samples.

Since weighted average treatment effects under confounding and endogenous continuous treatments can be regarded as an example of the WAD in a NPIV model, it is important to conduct some detailed Monte Carlo studies to compare finite-sample performance of various efficient estimators of $\theta_0$ when $h_0(Y_2)$ depends on multi-dimensional covariates $Y_2$.
In this paper we present large scale simulation studies focusing on the performance of several estimators of $\theta_0$ when $h_0(Y_2)$ is approximated via various ANN sieves and $Y_2$ is up to $13$-dimensional vector of continuous covariates.

\subsection{Efficient score and efficient variance for $\theta$}

In this section, we specialize the general efficiency bound result of AC12 to our setting.
We rewrite our model using their notation. Denote the full parameter vector as $\alpha_0 \equiv
(\theta_0 ,h_0)\in \Theta \times \mathcal{H}\equiv \mathcal{A}$. The model can
be written as the following sequential moment restriction
\begin{eqnarray}\label{ac}
    \E[\rho _{2}(Z, h_0(\cdot ))\mid X]& \equiv & \E[Y_1 - h_0(Y_2) \mid X] = 0,\quad
    \text{$X$ a.s.}  \\
    \E[\rho_{1}(Z,\alpha_0 )] &\equiv& \E[a(Y_2)\nabla_1 h_0(Y_2)-\theta_0 ]=0
    \nonumber
\end{eqnarray}
We define the orthogonalized residual as 
\[\varepsilon_1(Z,
\alpha) \equiv
\rho_1 (Z, \alpha) -\Gamma(X)\rho_2(Z,h) = a(Y_2) \nabla_1 h(Y_2) - \theta - \Gamma(X) \cdot (Y_1 -
h(Y_2)),\]
which is the residual from a projection of $\rho_1$ on $\rho_2$ conditional
on $X$, where $ \Gamma(X)$ is the orthogonal projection coefficient:
\[
 \Gamma(X)\equiv \frac{\cov(\rho_1(Z,\alpha_0) \rho_2(Z,h_0) \mid X)}{\var(\rho_2 (Z, h_0) \mid
X)}.
\]
Orthogonalizing the two moment conditions makes an efficiency analysis
tractable---the same technique is used in, e.g., \cite{chamberlain1992jbes}.

We now specialize the results in AC12 to the
plug-in model:%
\begin{equation}
\E[\rho _{2}(Z,h_{0}(\cdot )) \mid X]=0\text{ and }\E[\varepsilon _{1}(Z,\alpha
_{0})]=0,  \label{plugin}
\end{equation}%
where $\theta $ is a scalar and $h$ is a real-valued function of $Y_2$, and $\alpha = (\theta,h)$.
Define the following variances: 
\begin{equation*}
\sigma^2 _{0}\equiv \E[\{\varepsilon _{1}(Z,\alpha _{0})\}^{2}]=\var
\bk{a(Y_2) \nabla_1 h_0 (Y_2) - \theta_0 - \Gamma(X) (Y_1 - h_0(Y_2))}
\end{equation*}
\begin{equation*}
\Sigma(X)\equiv \var(\rho_2 (Z, h_0) \mid
X)=\var(Y_1- h_0(Y_2)) \mid X).
\end{equation*}

We recall the efficiency bound characterization for WAD of a NPIV model from AC12 (see their Example 3.3) for the sake of easy reference, and compute
\begin{equation}
J_{0}\equiv \inf_{r\in \overline{\mathcal{W}}}E\left\{ \{\sigma _{0}\}^{-2}%
\left(1+\E
    [a(Y_2)\nabla_1 r(Y_2) + \Gamma(X) r(Y_2)]\right) ^{2}+\Sigma(X)^{-1} \pr{\E[r(Y_2) \mid X]}^2\right\} 
\label{w8}
\end{equation}%
where $\overline{\mathcal{W}}=\{r:\E[\Sigma(X)^{-1}(\E\{r(Y_{2})|X\})^{2}]+\left(
E\{a(Y_{2})\nabla_1 r(Y_{2})+ \Gamma(X) r(Y_2)\}\right) ^{2}<\infty \}$. Let $r_{0}\in
\overline{\mathcal{W}}$ be one solution (not necessarily unique) to the optimization
problem (\ref{w8}). We note that such a solution always exists since the problem is
convex, and we have:%
\begin{equation}
J_{0}=\frac{1+\E[a(Y_2)\nabla_1 r_0(Y_2) + \Gamma(X) r_0(Y_2)]}{\sigma _{0}^{2}}
\label{J0}
\end{equation}

\begin{remark}[Characterization of Efficient Score]
\label{thm:eff-ai-chen-thm2.3} 
Applying Theorem 2.3
of AC12, we have: the semiparametric efficient score
$S^{\ast }$ for $\theta_{0}$ in \eqref{plugin} is
given by
\begin{equation*}
S^{\ast }(Z) =\frac{1+\E
    [a(Y_2)\nabla_1 r_0(Y_2) + \Gamma(X) r_0(Y_2)]}{\sigma _{0}^{2}}\varepsilon _{1}(Z,\alpha _{0})
    + \frac{\E[r_{0}(Y_2)|X]}{\Sigma (X)}(Y_1 - h_0 (Y_2))
\end{equation*}%
where $r_{0}\in \overline{\mathcal{W}}$ is one solution to \eqref{w8}. And the semiparametric information bound for $\theta _{0}$ is $%
J_{0}\equiv \var(S^{\ast })$.

(1) If $J_{0}=0$, then $\theta _{0}$ cannot be estimated at the $\sqrt{n}$-rate.

(2) If $J_{0}>0$, then the semiparametric efficient variance for $\theta
_{0}$ is: $\Omega _{0} \equiv (J_{0})^{-1}$.
\end{remark}

In the rest of the paper we shall assume that $J_{0}>0$ and hence $\theta_0$ is a
$\sqrt{n}$-estimable regular parameter. We note that by definition, the efficient score
(indeed any moment condition proportion to an influence function) automatically satisfies
the orthogonal moment condition.

\subsection{Efficient influence function equation based procedure}

From \cref{thm:eff-ai-chen-thm2.3}, the semiparametric efficient influence function for $\theta_0$ takes the form 
\begin{equation}
    \psi^*(Z,\theta_0) \equiv (J_{0})^{-1} S^*(Z) =\varepsilon _{1}(Z,\theta_{0},h_0)+(J_{0})^{-1}\frac{\E[r_{0}(Y_2)|X]}{\Sigma (X)} (Y_1 - h_0(Y_2)),
    \label{eq:eif}
\end{equation} 
Denote 
\[
\alpha_{e} (X)\equiv (J_{0})^{-1}\frac{\E[r_{0}(Y_2)|X]}{\Sigma (X)}.
\]
It is clear that $\theta_0$ is the unique solution to the efficient IF equation $\E[ \psi^*(Z,\theta_0)]=0$, that is
\[
\E\left[a(Y_2) \nabla_1 h_0(Y_2) - \theta - [\Gamma(X) -\alpha_{e} (X) ](Y_1 -
h_0(Y_2))\right]=0 \iff \theta=\theta_0.
\]
One efficient estimator, $\hat{\theta}_{ES}$, for $\theta_0$ is simply based on the sample version of the efficient IF equation with plug-in consistent estimates of all the nuisance functions:
\[
\hat{\theta}_{ES}=n^{-1}\sum_{i=1}^n\left( a(Y_{2i}) \nabla_1 \hat{h}(Y_{2i}) -[ \hat{\Gamma}(X_i) -\hat{\alpha}_{e}(X_i) ] (Y_{1i} -
\hat{h}(Y_{2i}))\right).
\]
In this paper $\hat{h}(Y_2)$ can be various ANN sieve minimum distance estimators (see below), but, for simplicity, the nuisance functions $\hat{\Gamma}(X)$ and $\hat{\alpha}_{e}(X) $ are estimated by plug-in linear sieves estimators.

\subsection{Optimally weighted SMD procedure}

Another efficient estimator for $\theta_0$ can be found by optimally-weighted sieve minimum
distance, where the population criterion is  (see AC12): 
\begin{equation} Q^0(\alpha) = 
     \E[m'(Z, \alpha) W_0(X) m(Z, \alpha)] = \E\bk{\frac{1}{\sigma^2_0}[\E
(\varepsilon_1
(Z,
\alpha))]^2 + \frac{1}{\Sigma(X)} (\E[Y_1 - h(Y_2) \mid X])^2}
\label{eq:smd_pop}
\end{equation}
The discrepancy measure is the optimally weighted quadratic distance of the expectation of the two moment conditions
\[
m(X, \alpha) = \colvecb{2}{\E[\varepsilon_1(Z, \alpha)]}{\E[Y_1 - h(Y_2)
\mid X]} = \colvecb{2}{\E[a(Y_2) \nabla_1 h (Y_2)-\theta - \Gamma(X) (Y_1 - h(Y_2))]}{\E[Y_1 - h(Y_2)
\mid X]} 
\]
from zero, where the optimal weight matrix $W_0(.)$ is diagonal and proportional to the inverse variance
of each moment condition:  \[
W_0(X) = \begin{bmatrix}
    1/\sigma_0^2 & 0 \\ 
    0 & 1/\Sigma(X) 
\end{bmatrix}
\]
Two remarks are in order. First, note that the optimal weight matrix $W_0(X)$ is diagonal
because $\varepsilon_1$ and $\rho_2$ are uncorrelated by design.  Second, since
the optimal weight matrix is diagonal and $\theta$ is a free parameter, we can view the
minimization as sequential: 
\[ h_0 = \argmin_{h\in\mathcal{H}} \E\bk{\frac{1}{\Sigma(X)}
(\E[Y_1 - h(Y_2) \mid X])^2},
\quad \theta_0 = \E[a(Y_2) \nabla_1 h_0 (Y_2) - \Gamma(X) (Y_1 - h_0(Y_2))].
\] 
This is important because solving the model sequentially while maintaining efficiency suggests a simple way to compute the estimators.

A sieve minimum distance estimator for $\alpha_0=(h_0,
\theta_0)$ may be constructed by (i) replacing expectations with sample means,
(ii) replacing conditional expectations with projection onto linear sieve bases,
(iii) replacing the optimal weight matrix with a consistent estimator, and (iv) replacing the
infinite dimensional optimization with finite dimensional optimization over a
sieve space for $h$. This paper focuses on approximating $h$ by ANN sieves. 
In particular, a sample analogue of the above objective function is 

\begin{equation*}
\widehat{Q}^0_{n}(\alpha )\equiv \frac{1}{n}\sum_{i=1}^{n}\widehat{m}%
(X_{i},\alpha )^{\prime }[\widehat{W }_0 (X_{i})]\widehat{m}%
(X_{i},\alpha )
\end{equation*} where $\widehat m(.;.)$ and $\widehat W_0(.)$ are estimators of
$m(\cdot, \cdot)$ and $W_0(\cdot)$ respectively; see \cref{sec:implementation,sec:MC}
below for examples of different estimators. Let $\mathcal{H}_n$ be a sieve parameter space
for $h$ (e.g., in this paper we focus on various ANN sieves). We define the optimally
weighted SMD estimator $\hat{\alpha}= (\hat{\theta},\hat{h})$ as an approximate solution
to
\[
\min_{ h\in\mathcal{H}_n,\theta \in \Theta} \widehat{Q}^0_{n}(h,\theta).
\]
This is an estimator proposed in AC12.

We may analyze the asymptotic properties of this estimator.
Since we may view the optimally weighted SMD problem as either a minimum distance program
or a sequential GMM estimator, we may carry out two separate analyses of the asymptotic
properties. The analysis of the estimator as a minimum distance problem is a
specialization of
\cite{ai2007estimation,ai2012semiparametric,ai2003efficient,CP2015sieve}, while
\cref{sub:cl} presents a heuristic review of the analysis as a sequential moment
restriction, which specializes \cite{CL2015gmm}. Either approach will lead to the
following asymptotic efficient influence function expansion:
\begin{equation}\label{eq:SMD-score}
\sqrt{n} (\hat \theta - \theta_0) = \frac{1}{\sqrt{n}} \sum_
{i=1}^n \bk{a(Y_2)\nabla_1 h_0(Y_{2i} )- 
\theta_0 - \pr{\Gamma(X_i) -
    \frac{\E[v_h^\star \mid X] }{\Sigma(X)}}  (Y_{1i} - h_0(Y_{2i}))} +
    o_p(1).
\end{equation}

\ \

\noindent {\bf Riesz Representer.} Lastly, we need to characterize the Riesz representer $v^\star$. The
argument in AC03 parametrizes $v^\star = v_\theta^\star (1, -w^\star)$ as a ``scale times
direction'' coordinate. For a fixed scale $v_\theta^\star$, the minimum norm property of
Riesz representers implicitly defines  the optimal direction $w^\star$ as the following:
\begin{equation}
    w^\star  =\argmin_w \E \bk{
    \frac{1}{\sigma_0^2} (\E[1 + a(Y_2)\nabla_1 w + \Gamma(X) w])^2 + \frac{1}
    {\Sigma(X)} (\E[w \mid X])^2
}.
\label{eq:optimalweightedwstar}
\end{equation}Solving the condition \[
\frac{1}{\sigma_0^2}\E[-v_\theta^\star +a(Y_2) \nabla_1
v_h^\star + \Gamma(X)
v_h^\star] = -1
\]
by plugging in $v_h = -w^\star v_\theta^\star$ then yields \[
v_\theta^\star = \frac{\sigma_0^2}{\E[1 +a(Y_2) \nabla_1 w^\star + \Gamma(X)
w^\star]}  \quad v_h^\star = \frac{-w^\star \sigma_0^2}{\E[1 +a(Y_2) \nabla_1 w^\star + \Gamma(X)
w^\star]}
\]
as the solutions for the representers where $w^\star$ is defined in 
\eqref{eq:optimalweightedwstar} above. If we assume completeness condition then $w^\star
 = r_0$ as the unique solution to \eqref{w8} or \eqref{J0} and $v_\theta^\star =(J_0)^
 {-1}$.

The consistency, root-$n$ asymptotic normality, consistent variance estimation can all be
obtained by directly applying AC03, AC07 for single hidden layer ANN sieves.
\cite*{chen2020nn2} results can be applied for multi-layer ANN sieves.

\section{Implementation of the estimators}
\label{sec:implementation}

In this section, we describe in broad strokes the implementation of the eventual
estimators for the average derivative of a NPIV, which often involves estimation of
nuisance parameters and functions. These nuisance parameters---which often take the form
of known transformations of conditional means and variances---require further choice of
estimation routines and tuning parameters, details of which are relegated to
\cref{sub:planned}.

\paragraph{A note on notation.} Recall that we use $Y_1$ to denote the outcome,
$Y_2$ to denote variables (endogenous or exogenous) that are included in the
structural function, and $X$ to denote exogenous variables that are excluded
from the structural function. Certain entries of $X$ and $Y_2$ may be shared.
Again, the NPIV model is:
\begin{equation}
    \E\bk{Y_1 - h_0(Y_2) \mid X} = 0.
\label{eq:npiv}
\end{equation}
Let $Z = [Y_1, Y_2, X]$ collect the observable
random variables (in the population). The parameter of interest is $\theta_0 = \E\bk{\nabla_1 h_0(Y_2)}$, where $\nabla_1
h_0(Y_2)$ is the partial derivative of $h_0$ with respect to its first argument,
evaluated at $Y_2$.

We also set up notation for objects related to the sample. Let there be a random
sample of $n$ observations. We denote $y_1 \in \R^n, y_2 \in
\R^{n \times p}, x \in \R^ {n\times q}$ as vectors and matrices respectively of
realized values of the random vector $(Y_1, Y_2, X)$. We will slightly abuse
notation and write $f(y_2)$, for a function $f : \R^p \to \R^d$, to be the $(n
\times d)$-matrix of outputs obtained by applying $f$ row-wise, and similarly
for expressions of the type $f(x)$.\footnote{This notation conforms with
how vector operations are broadcast in popular numerical software packages,
such as
Matlab and the Python scientific computing ecosystem (NumPy, SciPy,
PyTorch, etc.).} For a vector valued function $f$, we let
$P_f = f(x)(f(x)'f(x))^+ f(x)'$ be the projection matrix onto the column space
of $f(x)$.

\paragraph{Quick map of estimation procedures.} We provide a simple map that connects the above model and estimation 
approaches to the estimators we implement below.
\begin{enumerate}
    \item  (\cref{sub:smd})  For SMD estimators [\ismdns, \osmdns]: Solve sample and
    sieve version
    of \eqref{eq:smd_pop} 

    \item  (\cref{sub:if}) Score estimators [\isns, \esns]: Estimate the components of the
    influence functions as in (\ref{eq:eif}). Set the influence functions to zero and
    solve for $\theta$. 

    \item   (\cref{sub:inference}) Standard error for SMD estimators: Estimate the
    components of the
    influence functions as in (\ref{eq:SMD-score}), and take the sample variance.

\end{enumerate}
Additionally, we describe the estimator when the analyst is willing to
assume more semiparametric structure (e.g. partial linearity) on the
structural function $h_0(\cdot)$. We also conclude the section with a brief discussion
of software implementation issues.

\subsection{Sieve minimum distance (SMD) estimators}
\label{sub:smd}
Consider a linear sieve basis $\phi(\cdot)$ for $X$, where $\phi(X) \in
\R^k$. In sample, let $P_\phi$ be the projection matrix projecting onto the column space
of $\phi(x)$. 
For a sample of
realizations $v \in \R^n$ of $V$, let $P_\phi v$ be the sample best mean
square linear predictor (that approximates the conditional
mean) of $v$, since it returns the fitted values of a regression of $v$ on
flexible functions of $x$: \[P_\phi v \approx [\E[V_1 \mid
X_1],\ldots, \E[V_n \mid X_n]]'.\]
Under the NPIV restriction \eqref{eq:npiv}, taking $V = Y_1 - h_0(Y_2)$ and
$v = y_1 - h_0(y_2)$, we should expect \[
P_\phi (y_1 - h_0(y_2)) \approx 0.
\]
This motivates the analogue of the SMD criterion \eqref{eq:smd_pop} in the
sample, where we choose $h$ so as to minimize the size of the projected residual
$P_\phi(y_1
- h
(y_2))$:
\begin{equation}
    \hat h = \argmin_{h \in \mathcal H_n}\, \frac{1}{n}\norm[\big]{P_\phi [y_1 -
h
(y_2)]}^2.
\label{eq:smd_objective}
\end{equation}
When the norm chosen is the usual Euclidean norm $\norm{\cdot } = 
\norm{\cdot}_2$, we obtain the \textbf{identity-weighted SMD estimator} for $h_0$,
$\hat h_{\mathrm{ISMD}}$.

 Given a preliminary estimator $\tilde h$ for $h_0$, we may form an
estimator of the residual conditional variance $\Sigma(X) \equiv \E[
(Y_1-h_0(Y_2))^2 \mid X]$ by forming the estimated residuals $y_{1} -
\tilde h
(y_2)$  and then projecting $(y_{1} - \tilde h(y_2))^2$  onto $x$, e.g. via the linear sieve basis $\phi
(x)$ or via other nonparametric regression techniques such
as nearest neighbors. With such an estimator of the heteroskedasticity, we
can form a weight matrix $\hat W = \diag(\hat \Sigma(x))^{-1}$. Using the
norm $\norm{z}_W^2 \equiv z'Wz$ in \eqref{eq:smd_objective} yields the
\textbf{optimally-weighted SMD estimator} for $h_0$, $\hat h_{\mathrm{OSMD}}$.

With an estimated $\hat h$ of the structural function $h_0$, we can form
two plug-in estimators of $\theta$. The first is the \textbf{simple plug-in
estimator}: \[
\hat \theta_{\mathrm{SP}}(\hat h) = \frac{1}{n} \sum_{i=1}^n \nabla_1 \hat
h(y_
{2i}).
\]
See AC (2007) for the root-$n$ asymptotic normality of this estimator, and its asymptotic linear expansion is of the form:
\[
\sqrt{n} (\hat \theta_{\mathrm{SP}}(\hat h) - \theta) = \frac{1}{\sqrt{n}} \sum_
{i=1}^n \bk{\nabla_1 h(Y_{2i}) - 
\theta + \E[v_{h, \text{id}}^\star \mid X] (Y_{1i} - h(Y_{2i})) } +
    o_p(1).
\]
where \begin{equation}
    v_{h, \text{id}}^\star = \frac{-w^\star_{\text{id}}}{1 + \E[\nabla_1
w^\star_{\text{id}}]} \quad w^\star_{\text{id}} = \argmin_w \br{\E[\E
[w(Y_2) \mid X]^2] + (1 + \E[\nabla_1 w(Y_2)])^2}.
\label{eq:idw}
\end{equation} 
The simple plug-in estimator does not take into account the covariance
between the two moment conditions, $Y_1 - h(Y_2)$ and $\nabla_1(Y_2) -
\theta$. The second estimator, the \textbf{orthogonalized plug-in}
estimator, orthogonalizes the
second moment against the first:
\[
\hat \theta_{\mathrm{OP}}(\hat h, \hat \Gamma) = \frac{1}{n} \sum_{i=1}^n [\nabla_1 \hat
h(y_
{2i}) - \hat \Gamma(x_i) (y_{1i} - \hat h(y_{2i}))],
\]
where $\hat \Gamma$ is an estimator of the population projection coefficient of
the second moment $\nabla_1 h_0(Y_2) - \theta_0$ onto the first moment condition
$Y_1 - h_0(Y_2)$: \begin{equation}
    \Gamma(X) \equiv \E[(\nabla_1 h_0(Y_2) - \theta_0) (Y_1 - h_0(Y_2)) \mid X
] \Sigma^{-1}(X).
\label{eq:gamma}
\end{equation}
One choice of $\hat \Gamma$ is to plug in sample counterparts---plugging in
$\hat h$ for $h_0$, plugging in a preliminary $\hat \theta$ (which could be the
$\hat\theta_{\mathrm{SP}}(\hat h)$) for $\theta_0$, and plugging in an estimator
$\hat \Sigma$ for $\Sigma$---and finally approximate $\E[\cdot \mid X]$ via a
linear sieve regression, say with the basis $\phi(\cdot)$.

\ \ \

\noindent To summarize, the SMD estimator can be implemented as follows.

\ \ \

\noindent {\bf Identity Weighted SMD Estimator of $h(.)$}
\begin{enumerate}
    \item {\it Sieve for conditional expectation}: Choose a sieve basis $\phi(.)$ for $X$: $\phi(.) \in \R^k$ (more details on this later)
    \item {\it Construct objective function} \begin{enumerate} 
    \item 
    Obtain $P_\phi (y_1 - h(y_2))$ the sample least squares projection of $(y_1 - h(y_2))$ onto $\phi.$
\item {\it Optimizing $h(.)$:} define $\hat h = \argmin_{h \in \mathcal H_n}\, \frac{1}{n}\norm[\big]{P_\phi [y_1 -
h(y_2)]}^2_2.$ \end{enumerate}
\end{enumerate}

\noindent {\bf Optimal SMD Estimator of $h(.)$}
\begin{enumerate}
    \item Same as Step (1) above
    \item {\it Estimate weight function $\Sigma$}: with a preliminary estimator $\tilde h$ of $h$ (use identity-weighted one for instance), form an estimator $\hat \Sigma(x)$ by projecting $(y_1 - \tilde h(y_2))^2$ on $\phi(.)$, the sieve basis for $X$ to obtain $P_\phi((y_1 - \tilde h(y_2)^2)$. Form $\hat W = \diag(\hat \Sigma (x))^{-1}$. 
\item {\it Optimizing $h(.)$:} define $\hat h = \argmin_{h \in \mathcal H_n}\, \frac{1}{n}\norm[\big]{P_\phi [y_1 -
h(y_2)]}^2_{\hat W}$.
\end{enumerate}

\noindent {\bf Estimators for $\theta_0$}
\begin{enumerate}
    \item {\it Simple plug-in estimator.} Given an estimator $\hat h$ of $h$, use $$\hat \theta_{\mathrm{SP}}(\hat h) = \frac{1}{n} \sum_{i=1}^n \nabla_1 \hat h(y_{2i})$$
    \item {\it Orthogonalized plug-in estimator}
    \begin{enumerate}
        \item Obtain an estimator of $\Gamma$. One can use $\hat \Gamma(\hat \theta, \hat h) = P_\phi[(\nabla_1 \hat h(Y_2) - \hat \theta) (Y_1 - \hat h(Y_2))] \hat \Sigma^{-1}(X)$ with $\hat \theta$ being for example the simple plug in estimator and $\hat \Sigma(x)$ the above estimator of the variance of the first moment.
        \item Obtain $$ \hat \theta_{\mathrm{OP}}(\hat h, \hat \Gamma) = \frac{1}{n} \sum_{i=1}^n [\nabla_1 \hat h(y_{2i}) - \hat \Gamma(x_i) (y_{1i} - \hat h(y_{2i}))]$$
    \end{enumerate}
\end{enumerate}

Combining simple plug-in with identity-weighted SMD yields the estimation
procedure that we term \ismdns, and combining orthogonal plug-in with
optimally weighted SMD yields the estimation procedure that we call \osmdns.

\subsection{Influence function-based estimators} \label{sub:if}

We also implement influence function based estimators. 
As we highlighted in the previous section, one influence function estimator for $\theta_0$ takes the following form 

\begin{equation}
    \psi(Z, \theta, h, \kappa) = \nabla_1 h(Y_2) -
    \kappa(X)
    (Y_1 - h(Y_2)) - \theta.
    \label{eq:if}
\end{equation} with $\kappa(\cdot)$ defined below.
Moreover, 
given an estimator $\hat h$ for $h$ and $\hat \kappa$ for $\kappa$, we can
form the influence function estimator: 
\[
\hat\theta(\hat h, \hat \kappa) = \frac{1}{n}\sum_{i=1}^n \bk{\nabla_1 \hat h(y_{2i}) -
\hat\kappa(x_i)\pr{y_{1i} - \hat h(y_{2i})}}.
\]

\paragraph{Identity score estimator (\isns)} One influence function, which
corresponds to the influence function of the
\ismd estimator has $\kappa$ taking the following form. We refer to the
resulting influence function estimator as \isns, for \emph{identity score}.
\begin{align}
\kappa_{\mathrm{ID}}(X) &= \E[-v^\star(Y_2) \mid X] \\ 
v^\star(Y_2) &= \frac{-w^\star(Y_2)}{1 + \E[\nabla_1 w^\star(Y_2)]} \\ 
w^\star(Y_2) &= \argmin_{w} \br{
    \E\bk{\pr{\E[w(Y_2) \mid X]}^2} + \pr{1+\E[\nabla_1 w(Y_2)]}^2
}.\label{eq:w_star_id}
\end{align}

\paragraph{Efficient score estimator (\esns)}
On the other hand, the efficient influence function (\esns) uses a
different $\kappa(\cdot)$:
\begin{align*}
\kappa_{\mathrm{EIF}}(X) = \Gamma(X) - \E[v^\star(Y_2) \mid X]  \Sigma(X)^{-1} ,
\end{align*}
where $\Gamma(\cdot)$ is as in \eqref{eq:gamma}, and 
\begin{align}
v^\star(Y_2) &= \frac{-w^\star}{\E[1 + \nabla_1 w^\star + \Gamma(X) w^\star
(Y_2)]} \var
\bk{\nabla_1 h_0 - \theta_0 - \Gamma(X) (Y_1 - h_0(Y_2))}
\label{eq:v_star_optimal}\\
w^\star(Y_2) &= \argmin_w \br{
    \E\bk{\Sigma(X)^{-1} \pr{\E[w(Y_2) \mid X]}^2} + \pr{\frac{1+\E
    [\nabla_1 w(Y_2) + \Gamma(X) w(Y_2)]}{\sqrt{\var
\bk{\nabla_1 h_0 - \Gamma(X) (Y_1 - h_0(Y_2)) - \theta_0}}}}^2
}\label{eq:w_star_optimal}
\end{align}
are the same as \eqref{eq:optimalweightedwstar}, which are also weighted
analogues of the identity weighted $w^\star$, \eqref{eq:w_star_id}. 

The above formulation writes $v^\star$ as a function of
$w^\star$; alternatively, we may follow the strategy in \cref{sub:cl} and
estimate
$v^\star$ directly. One way to estimate
the above representer and hence get a feasible score is as follows.
Recall that the definition of $v^\star$ is \[
\E[\E[v^\star \mid X] \Sigma(X)^{-1} \E[v \mid X]] = \E[\nabla_1
v + \Gamma(X) v] \quad \norm{v^\star}_{\rho_2}^2 = \sup_v \frac{(\E
[\nabla_1
v + \Gamma(X) v])^2}{\E[ \Sigma(X)^{-1} \E[v \mid X]^2 ]}.
\]
Let $\nu(Y_2)$ be the basis approximating $Y_2$. Suppose we view that
$v^\star$
is well approximated by $\nu(Y_2)'\beta$, and that $\E[\cdot \mid X]$ is
well approximated by projection onto a basis $\lambda(x)$, then the above
definition of $v^\star$ yields a finite-dimensional problem that we may
solve in closed form to obtain the following. 
Consider the following quantities  \[
    F = \E[\nabla_1\nu(Y_2) + \Gamma(X) \nu(Y_2)]
   \text{ and }
    R = \E\bk{\Sigma(X)^{-1} \E[\nu \mid X]\E[\nu \mid X]'}.
    \]
    This then implies that $v^\star = \nu'R^{-} F$. In sample, this amounts
    to
    \begin{equation}
         \hat F = \frac{1}{n}\sum_i\bk{\nabla_1 \nu(y_{2i}) + \hat \Gamma(x_i)
    \nu(y_{2i})}
    \text{ and }
    \hat R = \frac{1}{n} \sum_{i}\bk{\hat \Sigma(x_i)^{-1}P_\lambda(x_i)
    \nu(y_
    {2i}) (P_\lambda(x_i) \nu(y_
    {2i}))'}
    \label{eq:vstar_cl}
    \end{equation}
   These can then be used to obtain $\hat v^\star$ and the influence
    function correction term \[
    \kappa_{\text{EIF}}(X) = \Gamma(X) - \E[v^\star(Y_2) \mid X]  \Sigma(X)^
    {-1}.
    \]

\ \ \

\subsection{Inference for \ismdns, \osmdns, \isns, \esns}
\label{sub:inference}

We now discuss how to compute standard errors and confidence intervals---again in broad
strokes---for the estimating algorithms \ismdns, \osmdns, \isns, and \esns. In a nutshell,
for the score estimators \is and \esns, the estimator $\hat\theta$ is a sample mean of
\emph{estimated} influence functions, and its sample variance is directly the properly
normalized variance of the influence functions. As a result, under appropriate conditions,
a sample variance of the estimated influence functions is consistent for the variance of
the influence functions, leading to consistent estimation of standard errors. For the
estimators \is and \esns, practitioners can therefore compute the standard errors without
adjusting for the estimation of the nuisance parameters.

Similarly, estimating the standard errors for the \ismd and \osmd estimators amounts to
estimating the variance of the influence function. One approach is to simply use the
influence function estimates from \is and
\esns, and leverage the fact that (\ismdns, \isns) and (\osmdns, \esns) are
respectively asymptotically equivalent. 

Another approach is to estimate the variance of the influence functions directly, without
necessarily estimating the influence functions themselves. The details are stated in
\cref{sec:model}, and we may turn the theory into estimators by ``putting hats on
parameters'': replacing unknown functions with their finite-dimensional sieve
approximations, conditional expectation with sieve projections, and expectations and
variance with their sample counterparts. For convenience, we reproduce the calculation
here:
\begin{enumerate}
    \item \ismdns: Consider \[
w^\star(Y_2) = \argmin_{w} \br{
    \E\bk{\pr{\E[w(Y_2) \mid X]}^2} + \pr{1+\E[\nabla_1 w(Y_2)]}^2
}
\]
which is the same as \eqref{eq:idw} and \eqref{eq:w_star_id}. Let $D_
{w^\star}(X) =
[-1-\E
[\nabla_1 w^\star], \E[w^\star \mid X]]'$. Then the asymptotic variance is \[
V = \frac{\E[\norm{D_{w^\star}(X)}^2]^2}{\E\bk{\norm{D_{w^\star}(X)}^2 (Y_1 - h_0
(Y_2))^2}}
\]
\item \osmdns: The inverse of the asymptotic variance is \[
V^{-1} = \min_w \br{
    \E\bk{\Sigma(X)^{-1} \pr{\E[w(Y_2) \mid X]}^2} + \pr{\frac{1+\E
    [\nabla_1 w(Y_2) + \Gamma(X) w(Y_2)]}{\sqrt{\var
\bk{\nabla_1 h_0 - \Gamma(X) (Y_1 - h_0(Y_2)) - \theta_0}}}}^2}
\]
which corresponds to the objective function in \eqref{eq:optimalweightedwstar}
\end{enumerate}

A third approach, which in our experience seems more accurate than
analytic standard errors, is a multiplier bootstrap for the SMD estimators.
The bootstrap simply replaces
the
residual $y_1 - h(y_2)$ in \eqref{eq:smd_objective} with the weighted
residuals $\omega(y_1 - h(y_2))$ where $\omega = \diag(\omega_1,\ldots,
\omega_n)$ are such
that $\omega_i \iid F_\omega$, independently of data, for some positively supported
distribution $F_\omega$ with unit mean and variance (e.g. the standard Exponential
distribution). Given a realization of the bootstrap weights $\omega$, the estimation
routines \ismd and \osmd would yield an estimate for $\theta$. Repeating this procedure
a large number of times would generate a large number of bootstrapped estimates, whose
percentiles form confidence interval boundaries.

\subsection{Partially linear or partially additive SMD estimators}
\label{sub:partiallylinear}
Assume $h_0$ is partially linear in its first argument, or, additionally,
partially
additive in subsets of its arguments. Since $h_0$ is linear in its first
argument, the slope on that argument is the average derivative $\theta_0$.
Therefore, under such a restriction, $h_0$ can be identified with the pair
$(\theta_0, \vartheta_0)$ where $\vartheta_0$ is some nuisance parameter governing
the rest of the function.

As in the case with SMD estimators in the nonparametric case, we solve the
SMD problem \eqref{eq:smd_objective}, while constraining $\mathcal H$ to
conform to the functional form assumptions made. The parameter $\theta_0$
is estimated via direct plug-in, since a solution $\hat h = (\hat\theta,
\hat \vartheta)$ for
\eqref{eq:smd_objective} naturally produces an estimator $\hat\theta$ for
$\theta_0$ \citep{ai2003efficient}.

\subsection{Implementation of neural networks}

We now provide a brief recipe on working with neural networks.
A feedforward neural network is a composition of \emph{layers} of the
form\footnote{For instance, a ReLU layer is a function of the form \[
x \mapsto \max(0, Wx + b)
\]
for $W$ a conformable matrix and $b$ a conformable vector.} \[
f_{\sigma, W, b} : \R^m \to \R^n \quad x \mapsto \sigma(W x + b) \quad 
\text{$\sigma : \R
\to \R$ is applied entry-wise.}
\]
for some conformable matrix $W$, vector $b$, and nonlinear 
\emph{activation function}  $\sigma$:, i.e. a $k$-hidden-layer neural
network has the
representation \[
h_\eta : \R^m \to \R^n  \quad h = b_{k+1} + W_{k+1} \cdot (f_{\sigma_k,
W_k, b_k}
\circ \cdots \circ
f_
{\sigma_1, W_1, b_1})
\]
where we collect the {learnable parameters} $\{W_j, b_j : j =1,\ldots, k+1\}$
as $\eta$. The gradient $\nabla_\eta h_\eta(y_2)$ can be computed
efficiently using the celebrated backpropagation algorithm, and, as a
result, in practice,  neural networks are often optimized via
first-order methods such as (stochastic) gradient descent or its variants, such
as the popular Adam algorithm \citep{kingma2014adam} in the machine learning
community. Optimization with neural networks is easiest with an
unconstrained, differentiable
objective, for which numerous computational frameworks exist. We use PyTorch \citep{paszke2017automatic} in this paper.\footnote{See \url{https://pytorch.org/}} In particular,
\eqref{eq:smd_objective} is an unconstrained, differentiable objective
function, and we may optimize over $\eta$ since the overall gradient may
be decomposed into components that are efficiently computed: By the chain
rule,\[
\nabla_\eta L(h, y_1, y_2, x) = \nabla_h L \cdot \nabla_\eta h,
\]
where $L(\cdot,\cdot,\cdot,\cdot)$ denote the objective
function \eqref{eq:smd_objective}.

Compared to conventional numerical linear algebra packages such as NumPy or
MATLAB, PyTorch offers two computational advantages particularly suited for deep
learning: automatic differentiation and GPU
integration. PyTorch tracks the history of computation steps taken to produce a
certain output, and automatically computes analytic gradients of the output with
respect to its inputs (See \cref{lst:torch} for an example).
Autodifferentiation allows gradient descent methods to be carried out
conveniently, without the user supplying analytical or numerical gradient
calculations manually. 

PyTorch also allows
arithmetic operations to be computed on GPUs, which have computing architecture
that allows for large-scale parallelization of simple operations. For
instance, multiplying two $k\times k$ matrices is of order $O(k^3)$ with a
naive algorithm, which can be viewed as $k^2$ dot products of size $k$; GPUs
allow for parallelized computing of the $k^2$ dot product operations, in
contrast to CPUs, where the level of parallelism is determined by the number of
CPU cores. For optimization, we use the Adam algorithm
\citep{kingma2014adam}, which is an enhancement of basic
gradient descent by estimating higher order gradients.
\begin{lstlisting}[language=Python,caption=Example of automatic
differentiation in PyTorch,label=lst:torch]
>>> import torch
>>> a = torch.tensor([1.], requires_grad=True)
>>> b = torch.tensor([2.], requires_grad=True)
>>> c = (a * b)
>>> c # we expect c = a * b = 2
tensor([2.], grad_fn=<MulBackward0>)
>>> c.backward() # Compute dc/da and dc/db
>>> a.grad # dc/da = 2
tensor([2.])
>>> b.grad # dc/db = 1
tensor([1.])
\end{lstlisting}

\subsection{Why linear sieves for certain nuisance parameters}
\label{sub:linearsieves}
\Copy{linearsieves}{
We note that even in our ANN implementation, a few nuisance functions are estimated with
linear sieves. For instance, the instrument projection, the conditional covariance
function $\hat \Gamma(X)$, and $w^\star$ in \eqref{eq:w_star_id} are all approximated
by linear sieves. It should be in principle possible to use nonlinear sieves, including
neural networks, for all of them, and we consider that to be important open work. Here,
we detail computational and conceptual difficulties that we have encountered.

First, many nuisance functions take the form of a conditional expectation $\E[r(Y_2)
\mid X]$, for some known or unknown function $r(\cdot)$. This is the case, for instance,
 with $\hat\Gamma$ in \eqref{eq:gamma}, as well as the conditional variance $\hat
 \Sigma$. In such cases, we can in principle use neural
 networks to minimize the empirical squared error loss\[
\min_{h_\eta \text{ is an ANN}} \frac{1}{n} \sum_{i=1}^n (\hat r(Y_{2i}) - h_\eta(X))^2
\] and use $\hat h_\eta$ as an estimate. At least computationally, such a procedure
 makes sense, though its theoretical properties may be delicate. In this work, we
 avoided using neural networks for $\hat \Gamma$ and $\hat \Sigma$ for computational
 convenience.

Replacing the instrument projection with neural networks is considerably more challenging
for sieve minimum distance.
In this case, we are not approximating a \emph{function}, so much as approximating an
\emph{operator} that projects onto $L^2(X)$. For a given estimate $\hat h$ with estimated
structural residuals $Y_{1} - \hat h(Y_2)$, it is not difficult to project it onto $X$ and
obtain the estimated projected residuals $\hat r (X;  \hat h)$ (by minimizing squared
error empirical risk), as well as its squared sample mean $\frac{1}{n} \sum_i \hat
r^2(X_i,
\hat h)$. However, it becomes challenging to update the neural network weights on
$\hat h$. Since the neural network $\hat r$ is \emph{trained} based on $Y_1 - \hat h
(Y_2)$, its weights depend on the weights of $\hat h$ in a complex fashion, and the
gradient of $\frac{1}{n} \sum_i \hat r^2(X_i,
\hat h)$ with respect to $h$'s weights become computationally intractable. As a result,
 we could not easily devise a scheme that replaces the instrument projection with
 nonlinear sieves. Recently, \cite{dikkala2020minimax} do propose a different method that
 allows for using nonlinear sieves for the instruments. The performance of this method is
 compared later in \cref{fig:f3id}. 

 Lastly, there are nuisance parameters which are defined through an optimization problem
 that includes its gradient with respect to its input. The nuisance parameters in the
 Riesz representer display this property, for instancec $w^\star$ in 
 \eqref{eq:w_star_id}. To approximate such a parameter, e.g. $w^\star$, with neural
 networks, we would minimize some criterion function that includes both $w$ and $\nabla_1
 w$. To use current off-the-shelf gradient-based training procedures, we would then
 require the gradient of $\nabla_1 w(\cdot)$ with respect to the $w$'s neural network
 weights, as well as that of $w(\cdot)$. The former is a niche use-case in deep learning,
 and so is not well-supported by the autodifferentiation methods in PyTorch. }

\ \ \

\section{Monte Carlo Studies}\label{sec:MC}

We present four Monte Carlo designs in the first subsection. We then describe exactly how
we estimated the various components that are needed for the estimators in the next subsection. The last subsection discusses some Monte Carlo results. 
\subsection{Design Descriptions}

We consider a set of Monte Carlo experiments that combine simple but relevant designs that
include high dimensional regressors. These designs are also relevant to the kinds of
empirical models that are of interests to economists. We describe the four Monte Carlo
designs below. A preponderance of our empirical results are based on \cref{mc:2}.

\ \ \ 

\begin{mc}
We thank an anonymous referee for suggesting these designs.\footnote{These designs
replace Monte Carlo 1 in an older version of the draft, which is in turn relegated to 
\cref{sub:amc}, now as \cref{mc:1}.} Part (a) of this
design
investigates performance of nonparametric regression (i.e. NPIV with the endogenous
variable equalling the instrument), as we vary the noise level. Part (b) of this design
investigates a simple NPIV design. Both designs feature neural networks as part of the
data-generating process, and as a result serve as optimistic benchmarks. 

    \label{mc:simple}
(a)  \Copy{condexp}{Consider the data-generating process \[
Y_i = f(X_i) + \sigma \epsilon_i \quad \epsilon_i \mid X_i \sim \Norm(0,1)
\]
where \[
f(x) = a_2'\tanh(A_1x+b_1) + b_2 \quad \text{ $\tanh$ applied entrywise}
\]
is a feed forward neural network with one hidden layer and 40 neurons. We consider
estimating $f (\cdot)$ for different $\sigma$ levels, calibrated to the variance of signal
$\var(f(X))$. The variance of the error is some multiple of the variance of $f$.\footnote{$
\var(f(X))$ is approximately 14 in our particular randomly generated $A_1, a_2, b_1,
b_2$.} We investigate four values of this multiple: 0, 0.1, 1, 10, corresponding to no
noise, moderate noise, high noise, and very high noise.}

(b)    
\Copy{mcsimple}{
    We generate i.i.d. standard Gaussians $W_i, Z_i$, where $W_i \in \R^p$ and $Z_i \in
    \R^2$. Let $X = (W', Z')'$.  We generate an endogenous treatment \[
R_2 = a_2'\tanh(A_1 X) + U_1 \quad U_1 \sim \Norm(0,1),
    \]
    for some fixed conformable coefficients $A_1, a_2$. Here $A_1$ has 4 rows, and
    $\tanh$ acts coordinate-wise. In other words, the endogenous   $R_2$   is a
    one-layer $\tanh$-network as a function of $W,Z$. 

    Let $U_2 = 0.9 U_1 + \sqrt{1-0.9^2} \Norm(0,1)$ be a normal correlated with $U_1$, and
    let \[
Y_1 = R_2 + a_3'W  + U_2
    \]
    be the second stage, for some fixed coefficients $a_3$. The coefficient of interest
    is on $R_2$, with its true value being $1$. To connect with the notation in the
    previous sections, let $Y_{2} = [R_1, W]$ and $X = [W, Z]$.
    }
\end{mc}

\ \ \

\begin{mc}
\label{mc:2}

The second Monte Carlo DGP is the following, which is an augmentation of the design in
\cite{chen2016methods}. 
\[
Y_1 =h_0(Y_2)+U= R_1  + h_{01}(R_2) + h_{02}(X_2)  + h_{03}(\tilde X) + U,
\quad \E[U \mid X_1, X_2, X_3, \tilde X] = 0, 
\]
where
\begin{align*}
h_{01} : \R \to \R &\quad t \mapsto\frac{1}{1+\exp(-t)}\\ 
h_{02}(t) : \R \to \R & \quad t \mapsto \log (1+t) \\
h_{03}: \R^{d_{\tilde x}} \to \R &\quad \tilde x \mapsto 5\tilde x_1^3 +
\tilde x_2 \cdot \max_{j=1,\ldots, d_{\tilde x}} \pr{\tilde
x_j\maxwith 0.5} + 0.5\exp(-\tilde x_{d_{\tilde x}}) \\
R_1 &= X_1 + 0.5U_2 + V \quad R_2 = \Phi(V_3 + 0.5 U_3) \\ 
 X_2 &\sim \Unif[0,1]\quad 
X_1 = \Phi(V_2) \quad X_3 = \Phi(V_3)\\ U &=
\frac{U_1+U_2+U_3}{3} \cdot \sigma(X_1, X_2, X_3) \\ 
\sigma(X_1, X_2, X_3) &= \sqrt{\frac{X_1^2 + X_2^2 + X_3^2}{3}}\\
U_\ell,V_k &\iid \Norm(0,1),\quad \ell = 1,2,3, k = 2,3 \\ V &\sim \Norm(0,
(\sqrt{0.1})^2).
\end{align*}

The process generating $\tilde X$ is
somewhat complex. First, we generate a covariance matrix $\Sigma \propto (I
+ Z'Z)$, normalized to unit diagonals, where $Z$'s entries are i.i.d.
standard Normal. The seed generating the covariance matrix is held fixed
over different samples, and so $\Sigma$ should be viewed as fixed a
priori. Next, let $\rho \in [-1,1]$ denote a
correlation level and we let \begin{equation}
\tilde X = \Phi\pr{\rho (X_1 + X_2 + X_3)  + \sqrt{1-\rho^2} T}\quad T \sim
\Norm(0,\Sigma),
    \label{eq:tildex}
\end{equation}
where $\Phi(\cdot)$ is the standard Normal CDF, and $\Phi(\cdot)$ and
addition are applied elementwise. In the exercises reported, we use $\rho
\in \{0, 0.5\}$ for correlation levels. In the high dimensional design, we set the
dimension of $\tilde X$ to be 10 and so the model will have 13 continuous regressors.

Note that this design allows for correlation among regressors both endogenous and
exogenous. It also allows for heteroskedasticity and possibly large dimensions by
increasing the dimension of $\tilde X$. We have also tried different conditional variance
of $U$, the simulation results are similar. To connect with the notation in the previous
sections, let $Y_{2} = [R_1, R_2, X_2, \tilde X]$ and $X = [X_1, X_2, X_3,
 \tilde X]$. The parameter of interest is $\theta_0 = \E\bk{\diff{h_0(Y_2)}{R_1}}=1$.

\end{mc}

\ \ \ 

\begin{mc}
\label{mc:3}

We modify \cref{mc:2} with two changes that allows for some nonlinearity of $h_0$ in $R_1$. In particular:
 
(a) $R_1$ enters $h_0(\cdot)$ through $R_1^2$. The parameter of interest is $\theta_0
= \E\bk{\diff{h_0(Y_2)}{R_1}}= \E[2R_1]=1$.

(b) $R_1$ enters $h_0(\cdot)$ through $R_1^2 / 2  + R_1 \frac{f(a(X_2 - b))}
{2C}$, where
\[
f(t) = h_{01}(t) (1-h_{01}(t)) \quad h_{01}(t) = \frac{1}{1+  e^{-t}}.
\]
and $C = \int_0^1 f(a(r-b))\, dr$, $a=-1$, and $b=16$. The parameter of interest is
\[
\theta_0 = \E\bk{\diff{h_0(Y_2)}{R_1}}=\E\bk{R_1 + \frac{f(a(X_2 - b))}{2C}} = \frac{1}{2} + \frac{1}{2} = 1.
\]

\end{mc}

Additionally, we provide a Monte Carlo calibrated to the empirical application in 
\cref{sub:gas}. 

\begin{mc}
\label{mc:calibrated}
\Copy{calibrated}{
    Like \cref{sub:gas}, we use 4,812 observations in the full sample as in 
    \cite{chen2018optimal}, taken from the 2001 National Household Travel Survey in 
    \cite{blundell2012measuring}. Each observation contains measurements of an outcome
    variable $y_1$ (log quantity of gasoline), an endogenous treatment variable $p$ (log
    price of gasoline), covariates $x$ (log income, log household size, log number of
    drivers in a household, log household age, total workers in household, and an
    indicator for public transit), and a price instrument $z$ (distance to the Gulf of
    Mexico).

    From a simple linear IV specification,\footnote{Regress $y$ on $p$ and log income,
    instrumenting for $p$ with $z$.} we estimate that the price elasticity of gasoline
    demand
    is $\epsilon_0 = -1.43$, and we create simulated data where $\epsilon_0$ is
    the ground truth. We do so by estimating the first stage relationship $\E[p \mid
    x,z]$ as well as the relationship of the outcome and the covariates $\E
    [y-\epsilon_0 p \mid x]$ nonparametrically, and build a simulation from these
    estimated quantities.

        \begin{enumerate}
        \item Estimate $f(x,z) = \E[p \mid x, z]$ with a single hidden layer (15
        neurons) sigmoid
        network. Estimate $g(x) = \E[y - \epsilon_0 p \mid x]$ with the same network
        architecture. To economize notation, we use $f(x,z), g(x)$ to denote the
        \emph{estimated} network rather than $\hat f, \hat g$. 
        \item Draw (with replacement) from the empirical distribution of the data and
        form $(Y_i, P_i, X_i, Z_i)_{i=1}^n$. For each variable $V$ in $(X,Z)$, we add
        Gaussian noise equal to
        10\% of the standard deviation of the variable $V$, in order to smooth the
        distribution of $V$. We use  notation $(X_i^*,
        Z_i^*)$ to denote the noised-up
        variables. 
        \item Let $R^P_i = P_i - f(X_i,Z_i)$ and $R^Y_i = Y_i - \epsilon_0 P_i - g(X_i)$
        denote the residuals for $f, g$. 
        \item Let $\hat P_i = f(X_i^*, Z_i^*)$ and $\hat Y_i = g(X_i^*)$ be the
        predicted
        price and residualized quantity from the noised-up synthetic data
        \item Let $P^*_i = \hat P_i + 1.3 R_i^P \cdot \eta_i   \equiv \hat P_i +
        \zeta_i $, $\eta_i \sim \Norm(0,1)$
        be the simulated price variable
        \item Let $\sigma(t) = \frac{1}{1+e^{-t}}$ denote the sigmoid function. Let
\[q(P^*_i, X_i^*) = \epsilon_0 \cdot \br{
\bk{\frac{g(X_i^*) - \mu_g}{\sigma_g} \cdot 0.2 + 1 - 5\bar{\sigma'}}\cdot P_i^* +
5\sigma
(P^*_i)} + g(X_i^*)
\]
where \begin{align*}
\mu_g =  \text{sample mean of } g(X_i^*) \quad \sigma_g = \text{ sample SD
of } g(X_i^*)
\\ \bar {\sigma'}=  \text{sample mean of }\sigma(P_i^*)(1-\sigma(P_i^*))
\end{align*}
The sample average derivative of $q(P_i^*, X_i)$ in $P_i$ is exactly $\epsilon_0$. As
the structural function of quantity in price (demand curve), it displays heterogeneous
price elasticities (in $X$) and nonlinearity in $P$. Generally speaking, the parameters
chosen ensure that the derivative of $q$ is negative. 
\item Let \[
\xi_i = 1.2 \zeta_i + R_i^Y \rho_i \quad \rho_i \sim \Norm(0,1)
\]
be the structural residual in the outcome, which is by design correlated with the
structural residual in the price DGP ($\zeta_i$).
\item Lastly, let $Y_i^* = q(P_i^*, X_i^*) + \xi_i$. 

    \end{enumerate}
The synthetic data is $(Y_i^*,
P_i^*, X_i^*, Z_i^*)$. To redraw the data, $(Y_i, P_i, X_i, Z_i, X_i^* - X_i, Z_i^* -
Z_i, \eta_i, \rho_i)$ are
redrawn, while $f, g$ are kept fixed.
}
\end{mc}

Next, we provide a step by step guidance on how to implement the estimators. 

\ \ \ 

\subsection{Implementation details} 
\label{sub:planned}
We explain here the exact choices of estimators that we used for these Monte Carlo
designs. A detailed overview is presented in \cref{tab:nuisance}. Various ANN SMD
estimators for $h$ have additional tuning parameters regarding nonlinear optimization,
which are described in \cref{tab:opt_tuning}.

Below, we describe the procedures underlying \cref{fig:f2}, which are representative of
the procedures in \cref{fig:r1,fig:mc3,fig:mc3_func}. We also describe the procedures for
\cref{fig:f3id,fig:f3op}, which are in turn representative of the procedures in
\cref{fig:es_sensitivity,fig:es_sigma_inv5000,fig:es_sigma_inv,fig:calibrated}.

\begin{table}[tb]
    
    \centering

    \begin{tabular}{ccc}
    \toprule
    Monte Carlo & Learning rate & \# steps \\ %
    \midrule
    1(b) & 0.01 & 6000--10000 \\
    2 & 0.01 & 3000--5000     \\  %
    3(a) & 0.01 & 7000--10000 \\
    3(b) & 0.01 & 7000--10000  \\ %
    4 & 0.005 & 6000--10000   \\
    5 & 0.001 &  1500--2000 \\ %
    \bottomrule
    \end{tabular}

    \caption{Optimizer parameter choices for ANN SMD for $h$ for the NPIV designs. The
    number of steps is of the form (minimum
    number of steps)--(maximum number of steps), where an ad hoc stopping rule is used
    when the step size is in between, based on how much progress the optimization
    procedure is making. In our experience, in practice, the optimizer stops at near the
    minimum number of steps. We use PyTorch's implementation of the Adam optimizer
    (\texttt{torch.optim.Adam}) throughout our experiments. }
    \label{tab:opt_tuning}
\end{table}
    
\begin{enumerate}

\item  \label{smd_est} \cref{fig:f2} reports Monte Carlo means and standard deviations for
the design in \cref{mc:2}, using ANN SMD estimators under a variety of
model specifications on true $h_0$. 

In particular, we make the following choices for estimation of various nuisance parameters.

    \begin{enumerate}
        \item Identity-weighted SMD with simple plug-in: $\hat \theta_
        {\mathrm{SP}}\pr{\hat h_{\mathrm{ISMD}}} $ defined in 
        \cref{sub:smd}. We specify choices of the linear sieve basis $\phi
        (\cdot)$ for instruments
        \begin{enumerate}
            \item $\phi(X) = [\phi_1(X_1,X_2,X_3), \phi_2(X, \tilde X)]$,
            where $\phi_1
            (X_1,X_2,X_3)$ follows the
            basis choice made
            in \cite{chen2007large} (p.5581--5582),\footnote{i.e. $
            \phi_1(X_1,X_2,X_3) = [1, X_1, X_1^2, X_1^3, X_1^4, 
            (X_1-0.5)_+^4,
            X_2, \ldots, X_2^4, (X_2-0.5)_+^4, X_3,\ldots, X_3^4, 
            (X_3-0.1)_+^4, (X_3-0.25)_+^4, (X_3-0.5)_+^4, (X_3-0.75)_+^4, 
            (X_3-0.9)_+^4, X_1X_3, X_2X_3, X_1 (X_3 - 0.25)_+^4, X_2
            (X_3-0.25)_+^4, X_1(X_3-0.75)_+^4, X_2(X_3-0.75)_+^4.]
            $, where $(\cdot)_+ = \max(\cdot, 0)$.} and $\phi_2(X, \tilde
            X) = [\tilde X, \tilde X^2, (X_i \tilde X_j)_{i,j}]$ contains
            second-order polynomials for $\tilde X$ and interactions $X_i
            \tilde X_j$. 
            
        \end{enumerate}
        \item \label{it:osmd} Optimally-weighted SMD with orthogonalized
        plug-in:
        $\hat \theta_{\mathrm{OP}}\pr{\hat h_{\mathrm{OSMD}}, \hat
        \Gamma}$ defined in 
        \cref{sub:smd}. We specify estimation details for the nuisance functions
        $\Sigma (X), \Gamma (X)$:
        \begin{enumerate}
            \item $\hat \Sigma(\cdot)$: Form the squared residuals from the
                    identity-weighted estimator $v \equiv (y_1 -
                                \hat h_{\mathrm{ISMD}}
            (y_2))^2$ and estimate $\Sigma$ by $k=5$-nearest neighbors.
            \item $\hat \Gamma(\cdot)$: Given an estimate $\hat
            \Sigma$, it suffices to estimate \[\E[(\nabla_1 h_0(Y_2) -
            \theta_0) (Y_1 - h_0(Y_2)) \mid X].
            \]
            Form $u \equiv \bk{\nabla_1 \hat h_{\mathrm{OSMD}}(y_2) -
            \hat
            \theta_
        {\mathrm{SP}}\pr{\hat h_{\mathrm{ISMD}}}} \pr{y_1 -
            \hat h_{\mathrm{OSMD}}
            (y_2)}$
            and project it on $\phi(X)$: i.e. $[\hat \Gamma(x_1),\ldots, \hat
            \Gamma(x_n)]' \equiv (P_\phi (\hat \Sigma^{-1} u))$.
        \end{enumerate}
    \end{enumerate}

For \cref{fig:f2}, 
    \begin{enumerate}
        \item The first column of \cref{fig:f2} reports results where the ANN
        SMD estimators are computed assuming $h_0$ is fully nonparametric.

            \item The second column of \cref{fig:f2} follows 
            \cref{sub:partiallylinear} in
    that we assume a partially linear structure on $h_0$, which is of the form $R_1 \theta + h
    (R_2, X_2, \tilde X)$.

    \item The third and the fourth columns of \cref{fig:f2} follow 
            \cref{sub:partiallylinear} in that we maintain the partially additive
             structure on $h_0$, which is of the form $R_1 \theta + h_1(R_2) + h_2
             (X_2) + h_3(\tilde X)$, where the unknown $h_3(\cdot)$ is approximated via
             ANNs. We use ANN sieves to approximate the scalar functions $h_1, h_2$ in
             the 3rd column, whereas the fourth column uses spline sieves to
             approximate $h_1, h_2$. 

    \end{enumerate}
    
    \item \label{score_est} \cref{fig:f3id,fig:f3op} reports Monte Carlo means and
    standard deviations for a wide class of estimators (not limited to
   ANN SMD) for \cref{mc:2}. 
    
    \begin{enumerate}
        \item ANN SMD: Follow \cref{smd_est} for \cref{fig:f2}.
        \item Spline SMD: 
        \begin{enumerate}
            \item \Copy{tensor}{Let $\lambda(x)$ be a spline basis for the
                    instrument space of $X$, and let $\nu(y_2)$ be a spline basis for the structural function $h(\cdot)$. 
                    Both $\lambda$ and $\nu$ are of the forms where each entry
                    expands into a $\text{Spline}(k,2)$ basis,\footnote{This notation is for
                    a spline with 2 knots, where, between adjacent knots, the spline
                    function is a polynomial of order $k-1$. } and pairwise interactions (of
                    the form $x_ix_j$, but we do not include more complex interactions $f(x_i) g
                    (x_j)$) are included in lieu of tensor product splines. The
                    choice of order $k$ for $\lambda(x)$ is 1 more than that for $\nu (y_2)$.}
        
         \item Given $\lambda, \nu$, we estimate P-ISMD, OP-OSMD as
         in \cref{smd_est}, where we optimize over candidate structural
         functions of the form
         $\nu(\cdot)'\gamma$, and estimate $\Sigma$ and $\Gamma$ by least squares projections onto the instrument sieve $\lambda$.
        \end{enumerate}
        
        \item Score/influence function estimators: Let $\lambda(x), \nu(y_2)$ be
        the spline bases used for the spline SMD in \cref{score_est}(b).
        \begin{enumerate}
            \item  IS: \begin{enumerate}
        \item Estimate $\hat h_{\mathrm{ISMD}}$ as in \cref{smd_est}(a) for ANN ISMD and as in \cref{score_est}(b) for spline ISMD.
            
        \item \Copy{wstar}{$v^\star(y_2)$ can be computed by solving 
        \eqref{eq:w_star_id}. To do so,
         we approximate $w^\star(y_2)$ with $\nu(y_2)\beta$ for some coefficients
         $\beta$, and the $\E[ \cdot \mid X]$ operator with $P_\lambda$. Doing so
         makes \eqref{eq:w_star_id} a least-squares problem in the unknown coefficients
         $\beta$. In fact, the closed form solution is \[
        \hat \beta = -\pr{\frac{1}{n}\nu(y_2)'P_\lambda \nu(y_2) + \frac{1}
        {n^2} \nabla_1 \nu(y_2)11' \nabla_1 \nu(y_2) }^{-1}\pr{\frac{1}{n}
        [\nabla_1 \nu(y_2)]'1},
        \]
        where $\nabla_1 \nu(y_2) \in \R^{n \times d_\nu}$ takes the partial
        derivative entry-wise. Therefore $\nu(y_2) \hat\beta$ is the
        estimator for $w^\star$, and this gives an estimator for $v^\star$ by
        plugging in.}
        \item Given $ \hat v^\star(y_2)$, we estimate $\kappa_{
            \mathrm{ID}}$ with $
        \hat \kappa_{\mathrm{ID}} (x) = P_\lambda \cdot \hat v^\star(y_2)$.
            \item Plug $\hat \kappa_{\mathrm{ID}}(x)$ and $\hat
            h$ to the inefficient influence function and compute $\hat\theta_{
            \mathrm{IS}}$.
        \end{enumerate}
        \item ES: \begin{enumerate}
            \item Estimate $\hat h, \hat \Gamma$ as in \cref{smd_est}(b) for ANN OSMD and as in \cref{score_est}(b) for spline OSMD.
            \item Estimate $v^\star$ by \eqref{eq:vstar_cl}. 
            \item Estimate $\Sigma$
            \item Form $\hat \kappa_{\mathrm{EIF}}(x) = \hat
            \Gamma(x) - P_\lambda [\hat v^\star(y_2)] \hat \Sigma(x)^
            {-1}$, where $\Sigma$ estimated via $k(n)$-nearest neighbors, with $k(n)>5$.
            \item Plug $\hat \kappa_{\mathrm{EIF}}(x)$ and $\hat
            h$ to the efficient influence function and compute $\hat\theta_{
            \mathrm{ES}}$.
        \end{enumerate}
        \end{enumerate}
                   \item AGMM: First we apply \cite{dikkala2020minimax}'s code to
                   estimate structural function $h_0$ by $\hat{h}_{\mathrm{AGMM}}$.  Then compute the simple plug-in  $\hat \theta_
        {\mathrm{SP}}\pr{\hat h_{\mathrm{AGMM}}} $ defined in 
        \cref{sub:smd}.       
    \end{enumerate}
        \end{enumerate}

\subsection{Monte Carlo Results}

Due to the length of the paper, we
report representative simulation results in a sequence of figures and tables
below.\footnote{\Copy{time}{As a note on computational difficulty, for a single run in
estimating
\osmd on a sample size of 5000 on a Mac
Mini (2020, Apple M1, 8GB RAM), the neural network procedures takes about 33 seconds for
\cref{mc:2}. Spline estimation usually takes about a second, as it is equivalent to
solving linear IV problems in closed form.} }

\subsubsection{Performance of point estimates in terms of (Monte Carlo) bias and variance}

\begin{figure}
    \includegraphics[width=\textwidth]{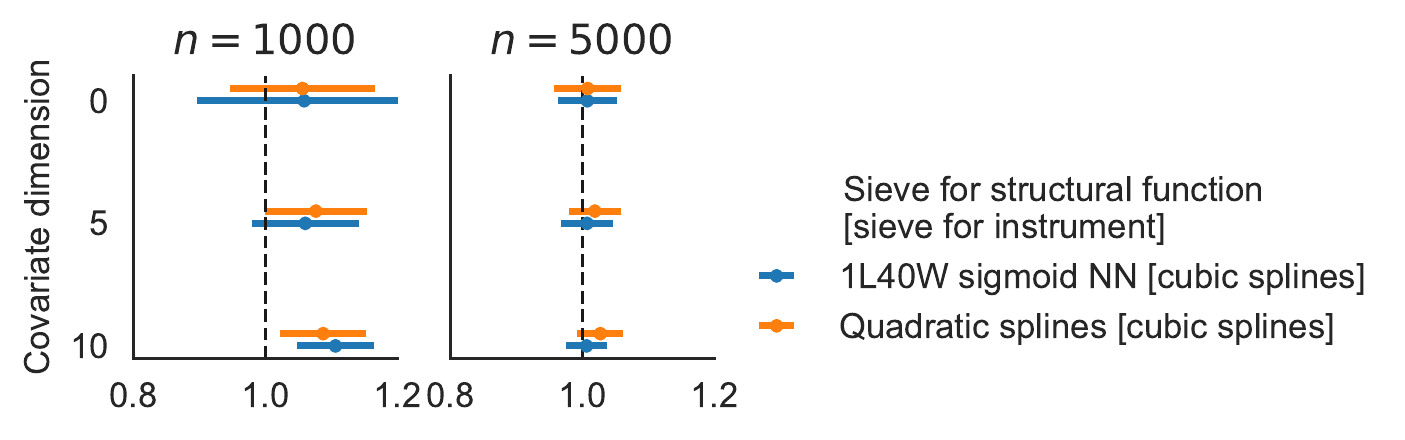}
    
    \caption{Performance of \ismd estimators on \cref{mc:simple}, where we vary over the
    dimension $p$ of the covariates $X$.}

    \label{fig:r1}
\end{figure}

\Copy{condexpresults}{For \cref{mc:simple}(a), we compare the performance of a ReLU
network (one hidden layer,
40 neurons) with the performance of a spline basis in \cref{table:cef}.\footnote{Like the
spline basis we use in other designs, it is a two-knot cubic spline for each variable with
pairwise variable interactions of the form $X_i X_j$.} The performance metric we choose is
the scaled integrated MSE: \[R^2 = 1-
\frac{\sum (\hat f(X_i) - f(X_i))^2}{\min_c \sum (f(X_i) - c)^2},
\] on 1000 out-of-sample data points. $R^2 = 1$ indicates perfect estimation of $f$, and
$R^2 = 0$ indicates estimation quality on par with using a constant prediction.
We find that for low and moderate noise, neural networks perform better than splines,
presumably since it captures more complex interaction patterns in $f$. In the high noise
regime, neural network underperforms, as neural networks overfit. In the very-high-noise
regime, both estimators overfit and are in fact worse than simply using a constant. We
caution that these performance of neural networks results from very minimal tuning, in
particular, without validation samples.}
\Copy{mcsimpledisc}{Next, for \cref{mc:simple}(b), we show the results in \cref{fig:r1}
for \ismdns{} estimators, varying over the dimension $p$. We see that, despite the DGP
involving a neural network, using
neural network estimators only attains a modest improvement over spline estimators, in
terms of slightly lower bias, consistent with our findings in the main text.} 

\begin{table}
\centering

\caption{Performance of ANN and spline for nonparametric regression}
\vspace{1em}

\label{table:cef}
\begin{tabular}{lrrr}
\toprule
  Noise-to-signal ratio & Spline $R^2$ & Neural net $R^2$ \\
\midrule
0.0 & 0.60 & 0.91 \\
0.1 & 0.58 & 0.80 \\
1.0 & 0.48 & 0.30 \\
10.0 & -0.55 & -2.55 \\
\bottomrule
\end{tabular}
\end{table}

The rest of the figures correspond to more difficult \cref{mc:2,mc:3,mc:calibrated}
where the first element of $Y_2$ is endogenous ($R_1$). %
\cref{fig:f2} reports the performance of various ANN SMD estimators for $\theta$
in \cref{mc:2}. The top display plots the results for $n=1000$ and the bottom
for $n=5000$. Note here that the columns correspond to various assumptions we
maintain on what the econometrician knows about the true structure of $h_0(.)$
in the model $\E[Y_1 - h_0(Y_2)|X]=0. $ The true design is partially additive,
and the first column, \textsf{NP}, assumes that the econometrician has no
knowledge of the true structure. As we can see, across all implementations (the
rows), most of the ANN SMD estimators perform well, which indicates that ANNs
seem able to adapt to the unknown structure of $h_0$. The second column labeled
\textsf{PL} (for partially linear) assumes that $h_0 (Y_2)$ is partially linear
(i.e., $h_0(Y_2) = \theta R_1 + h(R_2, X_2, \tilde X)$) while the third column
labeled \textsf{PA} assumes the correct additive structure (i.e., $h_0(Y_2) =
\theta R_1 + h_1(R_2)+h_2(X_2)+h_3(\tilde X)$) in the Monte Carlo design is
known to the econometrician (but the functions $h_1, h_2, h_3$ within it are of
course not known). \textsf{PA} column corresponds to the case where we use ANN
sieves to learn all the unknown functions $h_1, h_2, h_3$ although $h_1,h_2$ are
functions of scalar random variable. Its performance slightly deteriorates as
compared to the \textsf{NP} and \textsf{PL} columns. Notice here that for
comparison, the last column for the \textsf{PA} case uses splines to approximate
the two scalar valued unknown functions $h_1$ and $h_2$ while $h_3$ is always
estimated via ANN (since it is of higher dimensions (at least when $\dim(\tilde
X) >0$). We see that the spline results are in line with the \textsf{PL} and
   \textsf{NP} results, and are adequate here.
   
In \cref{fig:mc3}, we report results for the various estimators for \cref{mc:3}, where
the unknown function $h_0$ is now nonlinear in the endogenous $R_1$ (the first element of
$Y_2$). In the top panel (a), we report results for the case with $R^2/2$ and panel (b)
reports results for the case where the unknown function is $ R_1^2/2 + R_1f (X_2)$, where
now the derivative depends on the regressor $X_2$ nonlinearly (as the function $f$ is
highly nonlinear). Both results are for $n=1000, 5000.$ For panel (a) we see that the
spline estimator remain well behaved across all designs (across rows), the single-hidden
layer (1L) sigmoid ANN estimators remain adequate while both versions of the AGMM
estimators \citep{dikkala2020minimax} exhibit some bias. In panel (b), spline remains
well behaved and so are the ANN estimators. In \cref{fig:mc3_func} we show estimates of
the partial derivative evaluated at various fixed values for some regressors. Though the
estimators do not track the function well, especially in the tails in the bottom display,
the average derivative is estimated well. Interestingly, 3L relu ANN\footnote{This fact
seems to be robust to architectural choices.} seems to estimate the derivative function
marginally better than splines, perhaps since ANNs are able to automatically generate
rich interaction behavior, whereas specifying tensor products for spline sieves is
somewhat onerous.

We now examine various implementation choices for \cref{mc:2}. In 
\cref{fig:f3id,fig:f3op}, we
compare various implementations of ANN
estimators and spline estimators in \cref{mc:2}. In \cref{fig:f3id}, we
compare identity-weighted estimators (\isns, \ismdns, \textsf{AGMM},
\isns-X). Note that 
\ismd and \osmd are the plug in and optimal plug in SMD estimators. In \cref{fig:f3op}, we compare optimally-weighted estimators
that are semiparametrically efficient under suitable regularity conditions
(\esns, \osmdns{}, \esns-X).\footnote{As a reminder, we consider the following
estimators: \is or identity weighted score estimator, \es or the efficient score
estimator, while \isns-X and \esns-X are score estimators with two-fold cross
fitting. } It is important at the outset to keep in mind that all ANN
implementations require some non-negligible tuning as the optimization problem
is non-convex and the problem itself with endogeneity, correlation among the
regressors, and high dimensions is not easy to tune. Also, currently and for
NPIV models, there is no theory for data driven approaches to picking width,
depth, or activation functions and finite sample behavior in our design varied
\citep*[For linear splines, there are data-driven choice of sieve terms,
see][]{chen2021adaptive}.\footnote{although we have not implemented any
data-driven choice of spline sieve terms in our paper.}  The results
across various combinations of $\dim(\tilde X)$ and correlations for
$n=1000,5000$ indicate first that ANN OP-OSMD and especially spline estimators
seem to behave best. In particular, spline estimators require little tuning and
are more stable than all ANN based estimators we use. The SMD ANN estimators are
adequate with slight bias for the single-layer, varying-width case. \is and
\es ANN estimators are generally less biased and slightly higher variance
than \ismd and \osmd ANN estimators, but we note that good performance of
\es (in the ANN case) is very sensitive to the choice of $\hat\Sigma(X)^{-1}$
in the ``optimally-weighted'' Riesz representer estimation. 
\cref{fig:es_sigma_inv,fig:es_sigma_inv5000} compare the performances of \es in a variety of choices for $\hat\Sigma(X)$. It is interesting that the poor choice of $\hat\Sigma(X)^{-1}$ leads to biased estimation of \es and its cross-fitted versions.

\Copy{calibrateddisc}{Lastly, \cref{fig:calibrated} examines various estimators on the
empirical calibration
\cref{mc:calibrated}. Consistent with the findings in other Monte Carlo settings, we
generally find that all estimators perform adequately, with similar performance across
a variety of neural architectures. We also continue to find that SMD estimators \ismd,
\osmd have slightly better mean-squared error performance than the score-based estimators
 in exchange for slightly higher bias. The one exception is \es with variance estimation
 with only five nearest neighbors, which performs best across the specifications. We
 conjecture that this is due to how we constructed the residuals in \cref{mc:calibrated}.
 In particular, we multiply standard Gaussians with estimated residuals $R_i^P, R_i^Y$,
 which may result in a conditional variance function that is highly non-smooth, and, as a
 result, a low number of nearest neighbor estimates performs well. In any case, the
 performance of \es and \esns{}-X continues to show that it is sensitive to $\hat
 \Sigma(X)$, as is shown in \cref{fig:es_sigma_inv,fig:es_sigma_inv5000}. }

See \cref{sub:amc} for additional Monte Carlo results.

\subsubsection{Performance of inference statistics}

\cref{tab:infe,tab:infe2} provides various inference statistics for the
ANN SMD estimators \ismd and \osmd for \cref{mc:2}, without assuming any
semiparametric structure on $h(\cdot)$ beyond smoothness. In particular, we
report bootstrapped confidence intervals for ReLU and sigmoid and for depths 1
and 3 when the dimension of the nuisance variables $\tilde{X}$ ranges from 0 to
10. The results are also given for sample sizes $n=1000$  and $n=5000.$ Across
all specifications, the two ANN estimators perform adequately.

In \cref{fig:f4}, we examine various standard error approaches for a set of
estimators in \cref{mc:2}. For each of these, we compute the MC standard
deviation, a feasible estimator based on the estimator variance derived from
theory, and a bootstrapped standard error. Overall, the theory and bootstrapped
standard errors are adequate. In unreported results, criterion (SMD) based
bootstrap confidence intervals showed reasonable coverage performance.

\subsubsection{Overall simulation findings}
\label{subsub:overallfindings}

Overall, it seems that ANN methods are useful in approximating potentially high
dimensional functions in NPIV models. Also, in the class of models we
investigated, choices of layers, widths or activation functions are not very
consequential in terms of finite sample performance. On the other hand, ANN
based estimators in these non-standard NPIV models are hard to tune, and a
researcher needs to choose many smoothing parameters. These ANN estimators are also
unstable in some runs as they are based on highly complex (and non-convex)
optimization programs. In addition, ANNs are not as effective in estimating
univariate functions. Finally, to our surprise, We find that various plug-in
spline SMD estimators appear stable, less biased generally and can outperform
ANNs for NPIV models even in high dimensional cases with 13 continuous
regressors.
    
\section{Empirical Illustrations}\label{sec:application}

We present two empirical applications of estimating average derivatives with respect to endogenous price of a nonparametric demand $h_0(Y_2)$ for some non-durable goods. We apply ANN sieves to approximate $h_0(\cdot)$ nonparametrically when its argument $Y_2$ consists of 7 covariates (for gasoline demand) and 6 covariates (for strawberry demand). In the existing literature researchers have used both data sets to estimate unknown $ h_0(\cdot)$ in the model $\E[Y_1 - h_0(Y_2) \mid X] = 0$ by assuming $h$ takes some parametric or semiparametric (such as partially linear) form to avoid the ``curse of dimensionality'' of $Y_2$. Although served as illustrations, our applications below are the first to estimate the endogenous demand function $ h_0(\cdot)$ fully nonparametrically when $\dim(Y_2)>5$.

\subsection{Gasoline demand}
\label{sub:gas}

We use data on gasoline demand from the 2001 National Household Travel Survey
\citep*{blundell2012measuring}. The sample we use
include 4,812 observations in the full sample as in
\cite{chen2018optimal}. We estimate an NPIV analogue of the model (11) in
\cite*{blundell2012measuring}, $\E[Y_1 - h_0(Y_2) \mid X] = 0$ where $Y_1$ is the log
 gasoline demand, and $Y_2$ is a vector of 7 random variables consisting of the log
 gasoline price (possibly endogenous) and the other included covariates following
 Column (3) in Table 2 of \cite*{blundell2012measuring}. The instrument is the distance
 from the Gulf coast. We define the estimand as the average price derivative of the
 unknown structural function $h_0(\cdot)$, which has an average elasticity
 interpretation. %
 {blundell2012measuring} via OLS, and

  \begin{table}[h]
  \caption{
 Estimates of price elasticity for gasoline in National Household Travel
 Survey data \citep*{blundell2012measuring}
}

\label{tab:blundell_table}
  \centering
  \vspace{1em}
\begin{tabular}{ccccc}
\toprule
        &P-ISMD & OP-OSMD & IS  \\ \midrule 
Sigmoid [1L] &-1.28  & -1.24 & -1.12  \\ 
 & [-1.69, -0.9] & [-1.64, -0.87] &  (0.22) \\ 
 Sigmoid [3L] & -1.24 & -1.28 & -1.11 \\ 
 & [-1.65, -0.9] & [-1.64, -0.87] & (0.22) \\ 
 ReLU [3L] & -1.27 & -1.25  & -1.14\\  
 & [-1.65, -0.9] & [-1.64, -0.87] & (0.22)\\
 \midrule 
Spline(3, 2) & -1.17 & -1.2& \\ 
& [-1.57, -0.8] & [-1.6,-0.8]\\ 
\midrule 
 & \cite{blundell2012measuring} OLS & OLS &  TSLS \\ 
  &-0.83 &-0.85 & -1.24   \\ 
 &(0.148) & (0.15) &  (0.2)   \\ 
  \bottomrule 
\end{tabular}

\begin{proof}[Notes]
The 7 included covariates ($Y_2$) are: log gasoline price, log income,
household size, driver, household age, number working, public transit
distance. We instrument gasoline price with distance to Gulf of Mexico.
\end{proof} 
  \end{table}
  
\Cref{tab:blundell_table} shows
our estimates for the average price elasticity (and bootstrapped $95\%$ confidence
intervals). Broadly speaking, these estimates
point to a similar range of values and are similar to a parametric two-stage
least-squares specification. Across estimator classes, the ANN SMD estimates are
slightly larger in magnitude than the spline SMD estimates and the ANN \is
estimates. Within the ANN SMD estimator class, architecture
choices of the networks do not appear to matter much for the result. 
  
\subsection{Strawberry demand}

  \begin{table}[h]
  \caption{Estimate of demand average derivatives from Nielsen strawberry
    demand data \citep{compiani2019market}}
    \label{tab:strawb}
        \begin{minipage}{.5\linewidth}
      \centering
      \medskip
      \textbf{\footnotesize Non-organic}
      \footnotesize \\ \medskip
       \begin{tabular}{c c c c} \hline \hline
                     & IS & P-ISMD & OP-OSMD\\ \hline\hline
             \multirow{2}{*}{Sigm [1L]} & -1.649 & -1.530 & -1.747 \\
             &\scriptsize(0.04) & \scriptsize(0.04) & \scriptsize(0.03) \\
             & & [-1.8, -1.7] & [-2.3, -1.8] \\
             \hline
             \multirow{3}{*}{Relu [1L]}     & -1.648   & -1.590 & -1.706 \\
             & \scriptsize(0.04)  &  \scriptsize(0.04)   & \scriptsize(0.04) \\
             && [-1.9, -1.7]  & [-2.3, -1.8] \\ \hline
             \multirow{3}{*}{Relu [3L]}  & -1.648  & -1.634 &  -1.659
              \\  & \scriptsize(0.04)  & \scriptsize(0.04)  & \scriptsize
              (0.06) \\ 
              && [-1.9, -1.55] & [-2.2, -1.5]
             \\\hline
             \multirow{2}{*}{Spline(3,2)} & -1.611 & -1.648 & -1.676 \\
             & \scriptsize(0.04) & \scriptsize(0.04) & \scriptsize(0.04) \\ \hline
             \hline
        \end{tabular}
    \end{minipage}%
    \begin{minipage}{.5\linewidth}
      \centering
      \medskip \textbf{\footnotesize Organic}
      \footnotesize \\ \medskip
       \begin{tabular}{c c c c} \hline \hline
                     & IS & P-ISMD & OP-OSMD\\ \hline\hline
             \multirow{2}{*}{Sigm [1L]} & -3.235  & -2.409 &  -3.382 \\
              & \scriptsize(0.07) & \scriptsize(0.09) & \scriptsize(0.06)
              \\
                           & & [-2.7, -2.44] & [-4.3, -3.5] 
              \\\hline
             \multirow{3}{*}{Relu [1L]}  & -3.236   & -2.197  &  -2.129 \\
             & \scriptsize(0.07)  & \scriptsize(0.06)   &  \scriptsize(0.08)\\
                           & & [-2.4, -2.11] & [-2.4, -2.06]
                            \\\hline
             \multirow{3}{*}{Relu [3L]}  & -3.232  & -2.206 &  -2.122
              \\
             & \scriptsize(0.07)  &  \scriptsize(0.07)    & \scriptsize(0.14) \\
                           & & [-3.1, -2.08] & [-2.36, -2.06]
                            \\\hline
             \multirow{2}{*}{Spline(3,2)} & -3.194 & -3.232 & -3.124 \\
             & \scriptsize(0.06) & \scriptsize(0.07) & \scriptsize(0.06) \\\hline
             \hline
        \end{tabular}   
    \end{minipage}
    \begin{proof}[Notes]{
         The 6 included covariates ($Y_2$) are: strawberry prices (non-organic,
organic), income, lettuce demand (taste for organic proxy), state-level
sale of non-strawberry fresh fruits, average outside good price.
        The excluded instruments are 3 Hausman IV 
        (prices in neighbouring markets)+ 2 strawberry spot prices 
        (marginal cost measures).  A market is defined at the store-week
        level and there are $N = 38,800$ markets. }
    \end{proof}
  \end{table}

We also consider a setting where consumers choose two substitutable
goods. We use the Nielsen dataset from \cite{compiani2019market},\footnote{Our results do not
necessarily represent the views of the Nielsen Company.} where consumers in
California choose from strawberries, organic strawberries, and an outside
option.\footnote{For a detailed description of the data, see Appendix G of
\url{https://www.tse-fr.eu/sites/default/files/TSE/documents/sem2019/eee/compiani.pdf}.}
We observe the market share of each type of product, their prices, and a variety
of covariates at the market (store-week) level. In the analysis, we consider
NPIV model $\E[Y_1 - h_0(Y_2) \mid X] = 0$ where $Y_1$ is the log market share
of a type of good (non-organic or organic strawberries) and $Y_2$ is a vector of
6 random variables, including endogenous prices for both types of strawberries
and the outside good, and other market-level covariates. The instruments $X$
include Hausman instruments as well as cost shifters such as measurements
of consumer taste and income at the market level.
We focus on the target parameter $\theta_0=\E[\nabla_1 h_0]$, which is the average
derivative of $h$ with respect to the own-price in logs, which we
interpret as a version of price elasticity.\footnote{Under a
model of the demand where the NPIV condition $\E[Y_1 - h_0(Y_2)
\mid X] = 0$ defines the demand function $h_0$, we can understand
$\theta_0$ as a price elasticity. However, this model---which implicitly assumes
that endogeneity is additive---may not be consistent with microfoundations of
consumer behavior \citep{berryhaile2016}, and so care should be taken in interpreting  $\theta_0$ as an elasticity. Nevertheless, for purposes of our illustration here,
we may continue to view $\theta_0$ as some well-defined function of the
distribution of the data. For a more detailed implementation of demand in this setting, see 
\cite{compiani2019market} where in principle one can also use the neural networks based implementation in this paper in a natural way.} 

We present the results in \cref{tab:strawb}. As is
perhaps expected from a casual intuition, estimates of $\theta_0$ are negative
across both products, and more negative for the more price-sensitive product
(organic strawberry). Moreover, results are broadly similar across estimation
methods (SMD vs. score) and sieve choices (spline vs. neural net), with perhaps
more variability for neural networks in organic strawberries.
The estimates for non-organic strawberries hover around $-1.5$, and are
reasonably stable across choices of tuning parameters and estimators 
(\is vs. SMD estimators). The estimates for organic strawberries are more
variable across specification of nuisance parameters and neural
architectures, but seem to be around $-2$ and $-3$, and larger in
magnitude than the own-price elasticity estimate for non-organic
strawberries. 

These estimates are qualitatively similar to 
\cite{compiani2019market}'s estimates, which reports median own-price elasticities of
 $-1.4$ (0.03) for non-organic strawberries and $-5.5$ (0.7) for organic
 strawberries.\footnote{Interestingly, our estimates are closer to estimates from BLP
 that \cite{compiani2019market} reports in Figure 4, which are also around -2 to -3.}
 Our estimates are more dissimilar for organic strawberries, for which we offer a few
 conjectures. First, 
\cite{compiani2019market} reports estimates following 
\cite{berryhaile2016}'s approach to demand estimation, that accounts for price endogeneity differently. Under his assumptions, it is possible that our estimator is consistent for a different parameter than his. Second, organic strawberry market shares
are very small, and hence fluctuates more on a log scale, thereby resulting in
worse estimation precision.

\section{Conclusion}

\label{sec:conclusion}

\Copy{conc}{
In this paper, we present two classes of semiparametric efficient estimators for weighted
average derivatives (WADs) of nonparametric instrumental variables regressions (NPIV) of
moderate and high dimensional endogenous and exogenous regressors. We have conducted
detailed Monte Carlo comparisons of finite sample performance of various inefficient and
efficient estimators of the WADs using various ANN sieves. The simulation studies and
empirical applications confirm the theoretical advantage of ANN approximation of unknown
continuous functions of moderately high-dimensional variables, after some tuning of
hyper-parameters. Perhaps the most practical findings from our large amount of reported
and unreported simulation studies using moderate sample sizes are as follows:
the ANN efficient SMD estimators have smaller biases than those of the ANN inefficient SMD
estimators, and are less sensitive to the tuning parameters than those of the ANN
efficient score estimators. 
 In addition, simple spline based estimators of WADs of NPIVs perform very well in terms of
finite sample biases and variances. More research is needed to close the gap between
approximation theory and finite sample computational performance in applying flexible ANNs
to nonparametric models with endogeneity. 
}
\newpage

\bibliographystyle{ecca}
\bibliography{QAD-biblio}

\newpage

\appendix

\newpage

\begin{figure}[htbp]
 \caption{ANN SMD estimators for \cref{mc:2}}
  \centering
  
  (a) $n = 1000$
  
  \includegraphics[width=\textwidth]{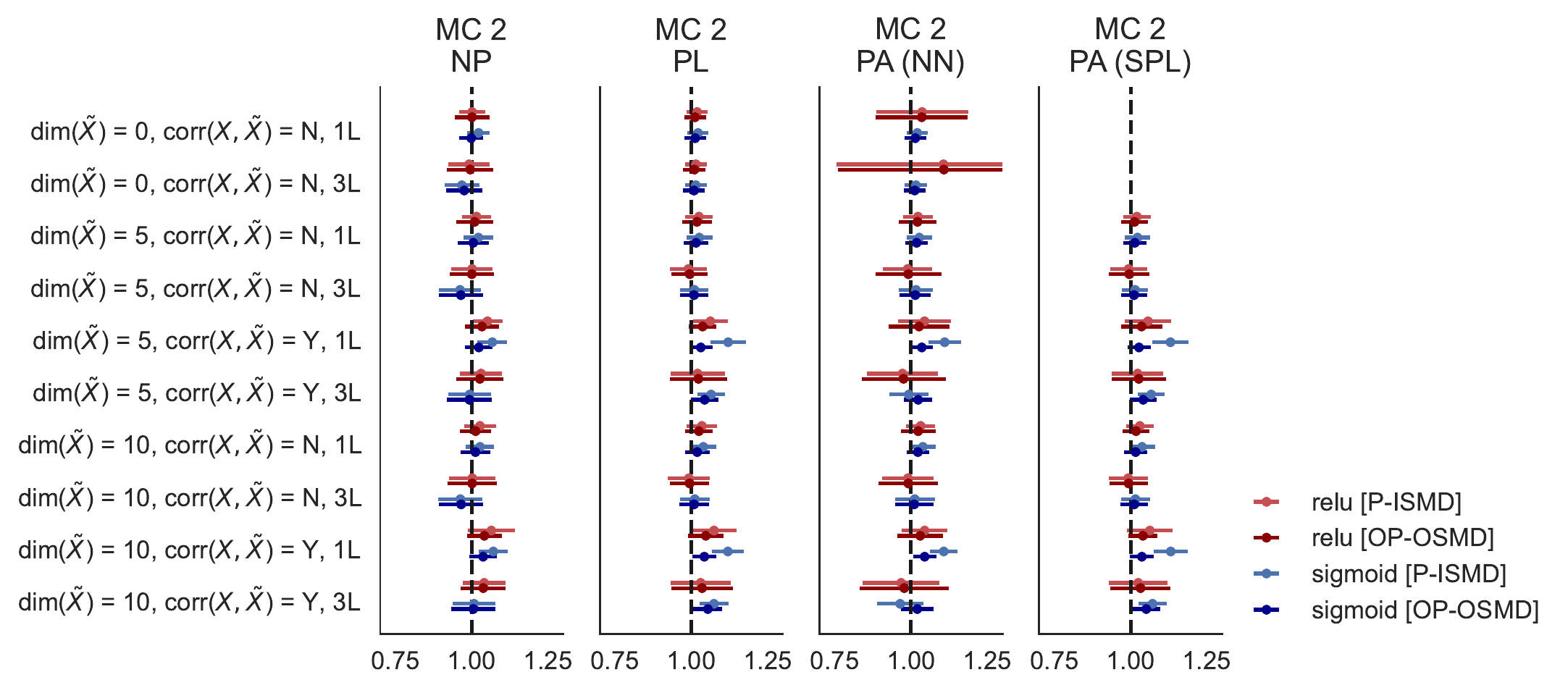}

  (b) $n = 5000$
  
  \includegraphics[width=\textwidth]{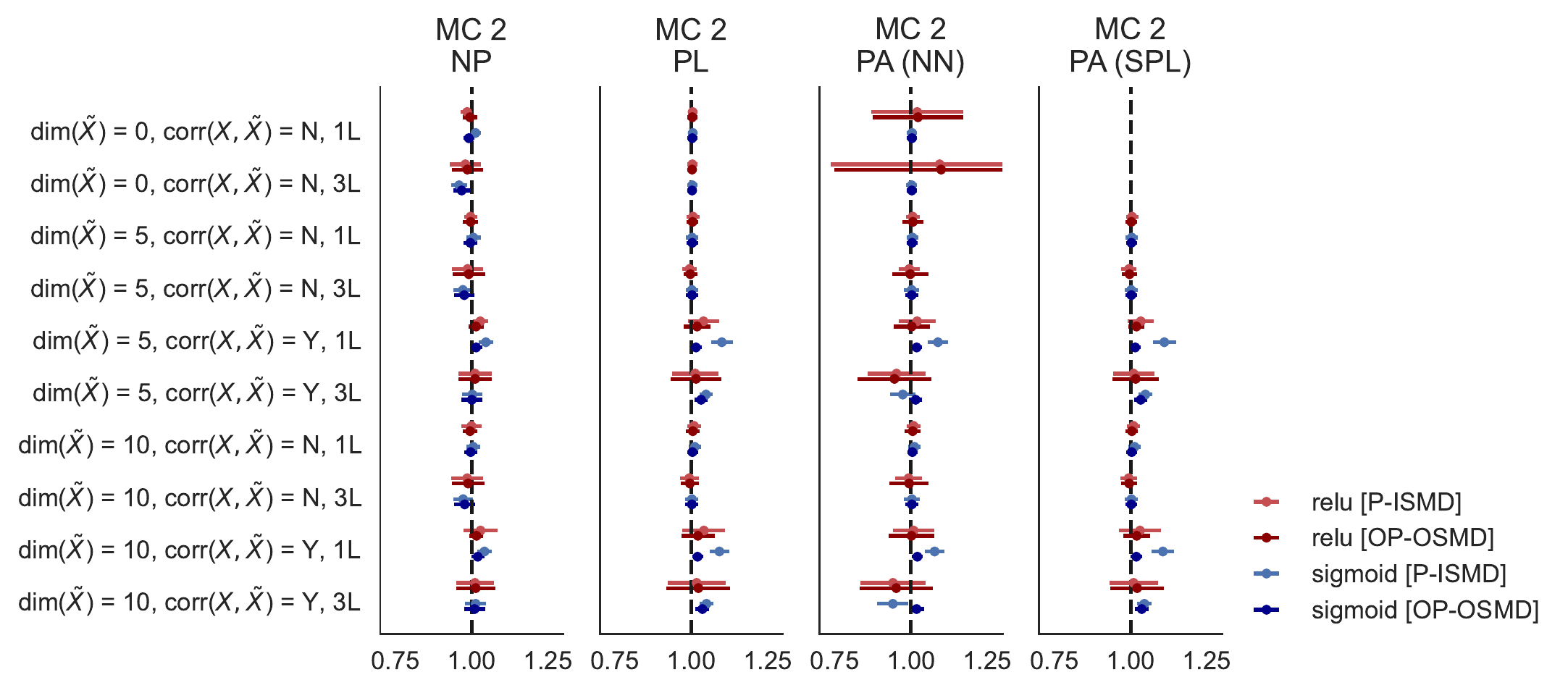}
  
  \begin{proof}[Notes]{\footnotesize
     Monte Carlo Mean $\pm 1$ Monte Carlo standard deviation across 1,000
replications. 

The columns are estimators where different correct assumptions of
the data-generating process are placed. The first column (NP: 
\emph{nonparametric}) shows estimated average derivative of an NPIV model,
where the unknown function $h(Y_2)$ is not assumed to have separable
structure. The second
column (PL: \emph{partially linear}) assumes $h(Y_2) = \theta R_1 + h_1
(R_2, X_2, \tilde X)$. The third and fourth columns (PA: \emph{partially
additive}) assumes $h(Y_2)= \theta R_1 + h_1(R_2) + h_2(X_2) + h_3(\tilde
X)$. The third column uses neural networks to approximate the scalar
functions $h_1, h_2$, and the fourth column uses splines to approximate
$h_1, h_2$ (while $h_3$ is always estimated via ANN).

For each type of assumption placed on the true $h_0(Y_2)$, we vary the
data-generating process by varying the dimension of $\tilde X$ and the
level of
correlation between $(X_1,X_2,X_3)$ and $\tilde X$. We also vary the
network architecture by $\{\text{ReLU}, \text{Sigmoid}\} \times \{
\text{1L, 3L}\} \times \{\text{10W}\}.$ Lastly, we vary the type of
estimator used from simple
plug-in with the identity-weighted SMD estimator to orthogonalized plug-in
with the optimally-weighted SMD estimator.}
 \end{proof} 
 
  \label{fig:f2}
\end{figure}

\begin{figure}[htb]
  \centering
  \caption{Estimation quality of average derivative parameter in 
  \cref{mc:3} across a variety of NPIV estimators} 
  \label{fig:mc3}
  \ \ 
  
  (a) \cref{mc:3}(a)
  \includegraphics[width=\textwidth]{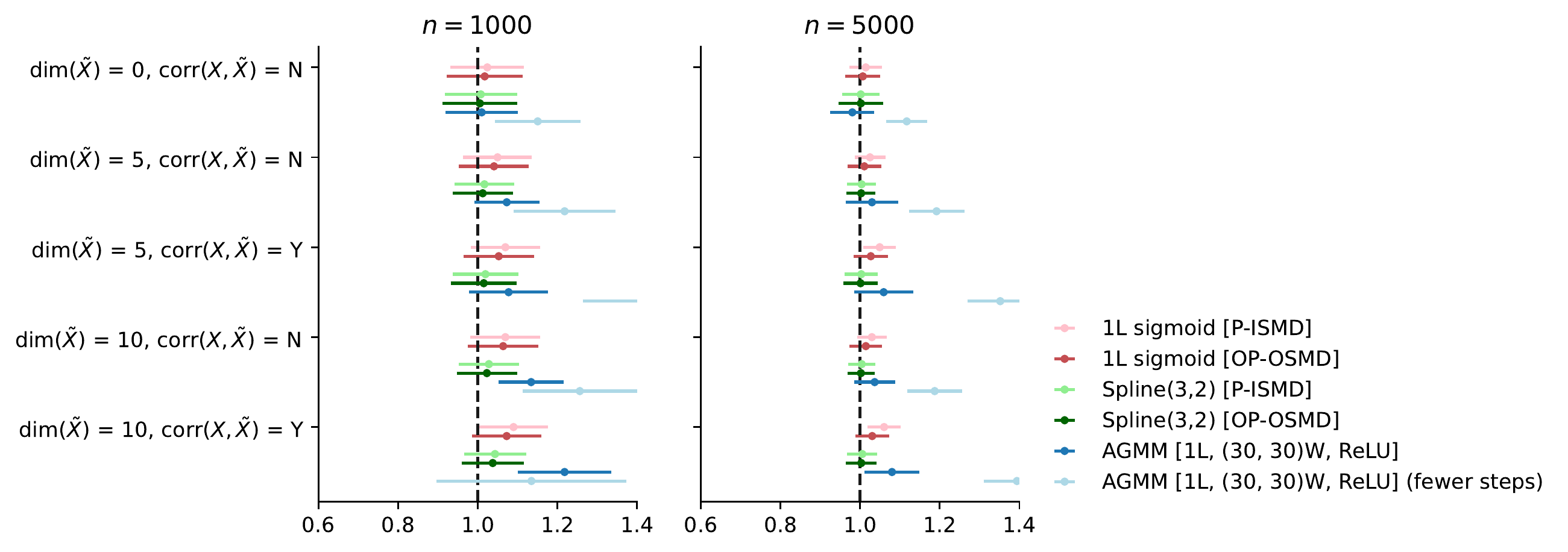}

\ \ 

    (b) \cref{mc:3}(b)
  \includegraphics[width=\textwidth]{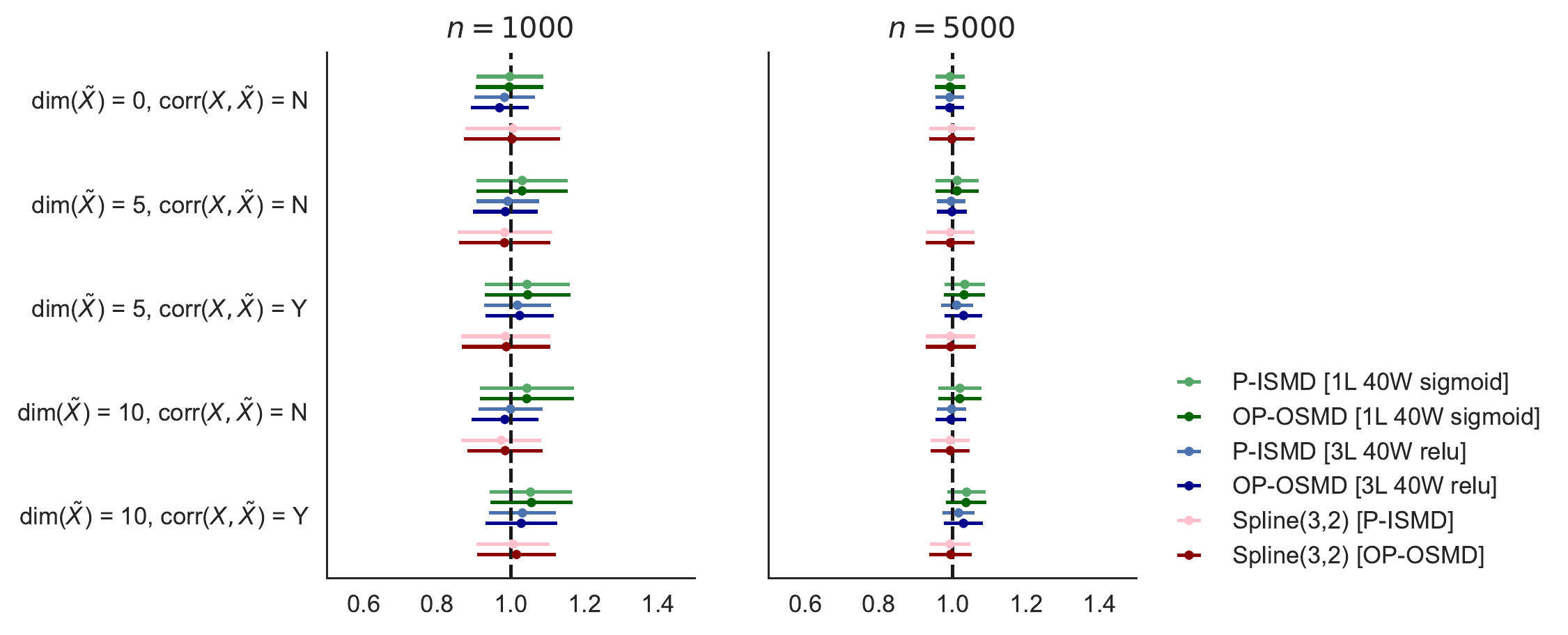}
  
\end{figure}

  \begin{figure}[htb]
  
  \caption{Estimation quality of the partial derivative function
  in
  \cref{mc:3}(b) across a variety of estimators}
\label{fig:mc3_func}
  \centering
    
    \ \ 
  (a) Estimated $f_1$ versus true $f_1$. Single sample for $N =
  5,000$

  \includegraphics[width=0.7\textwidth]{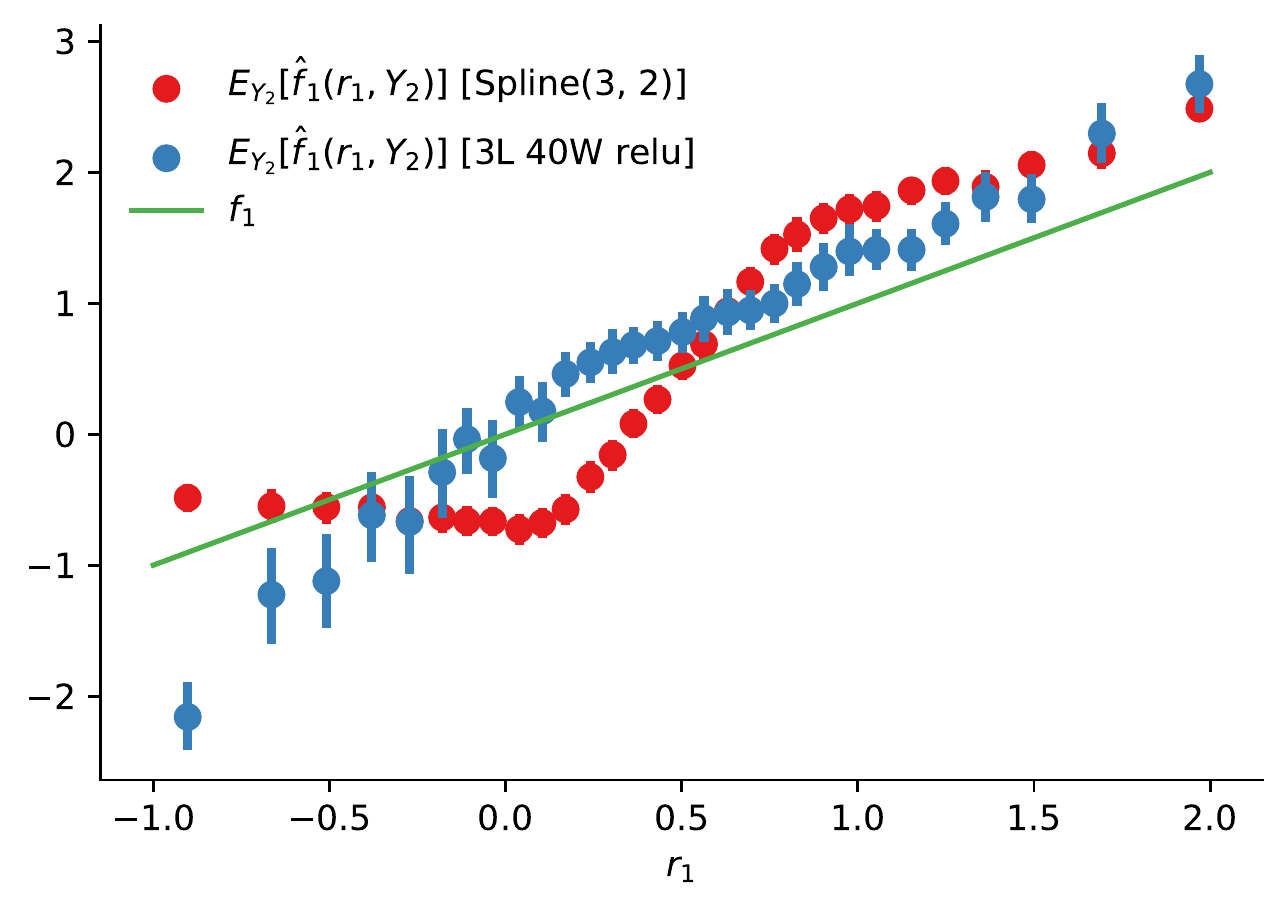}

(b) Estimated $f_2$ versus true $f_2$. Single sample for $N =
  5,000$

  \includegraphics[width=0.7\textwidth]{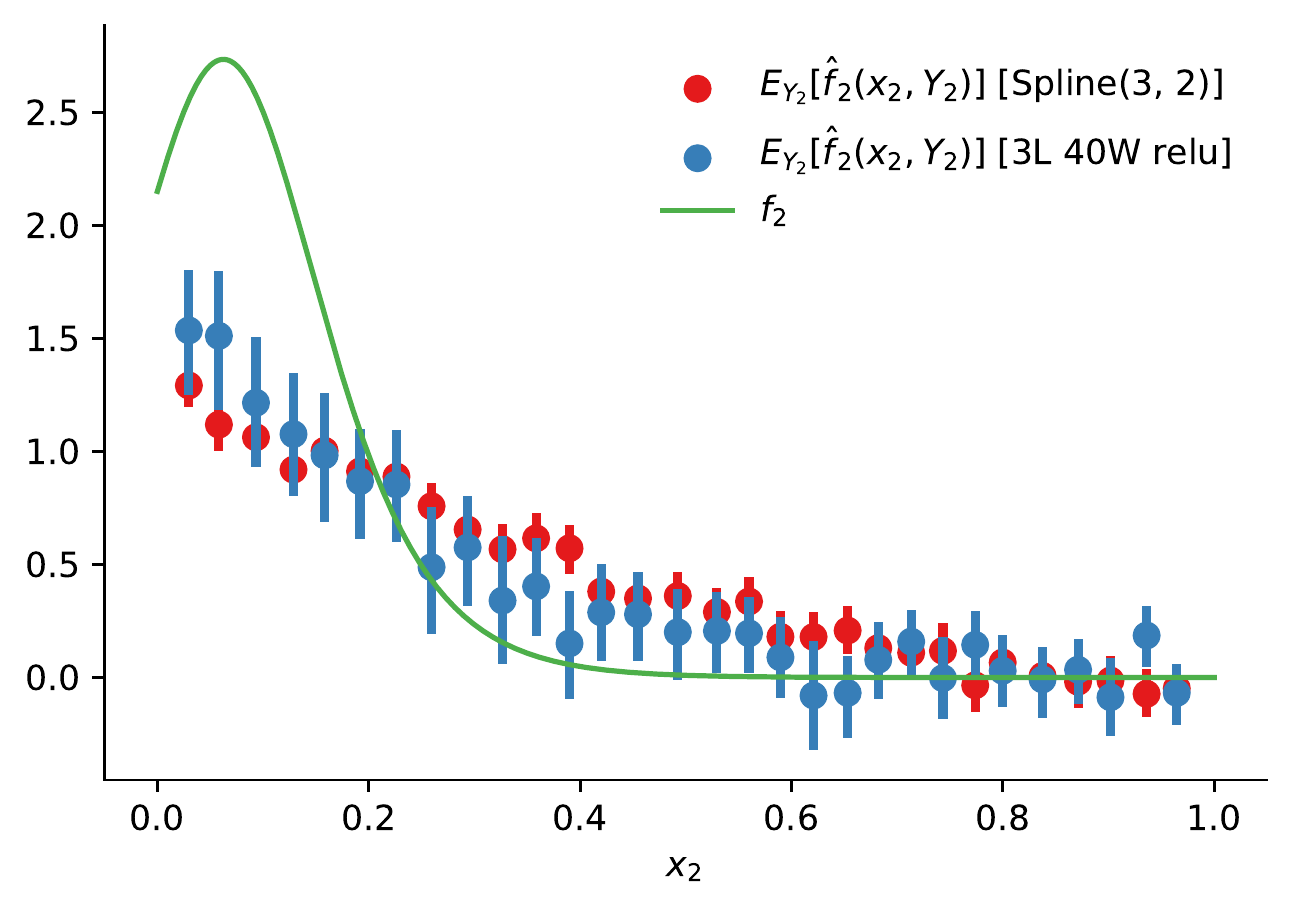}

           \begin{proof}[Notes]{\footnotesize
In the DGP \cref{mc:3}(b), the partial derivative $\nabla_1 h_0$ is of the form
$f_1 (R_2) + f_2(X_2)$, and we evaluate performance estimating $f_1,f_2$.
Estimated $f_1$ is calculated by taking $\nabla_1 \hat h - f_2(x_2)$. We plot
expectation marginalizing over variables other than $r_1$. Estimated $f_2$ is
calculated by taking $\nabla_1 \hat h - f_1(r_1)$. We plot expectation
marginalizing over variables other than $x_2$. Both estimators use the same instrument
basis.}
  \end{proof}

\end{figure}

  \begin{figure}[htb]
  
  \caption{Estimation quality of average derivative parameter in 
  \cref{mc:2} across a variety of \emph{identity-weighted} estimators}
  \centering
  \ \ 
  
    (a)  \cref{mc:2}, Nonparametric, $n=1000$
    
  \includegraphics[width=\textwidth]{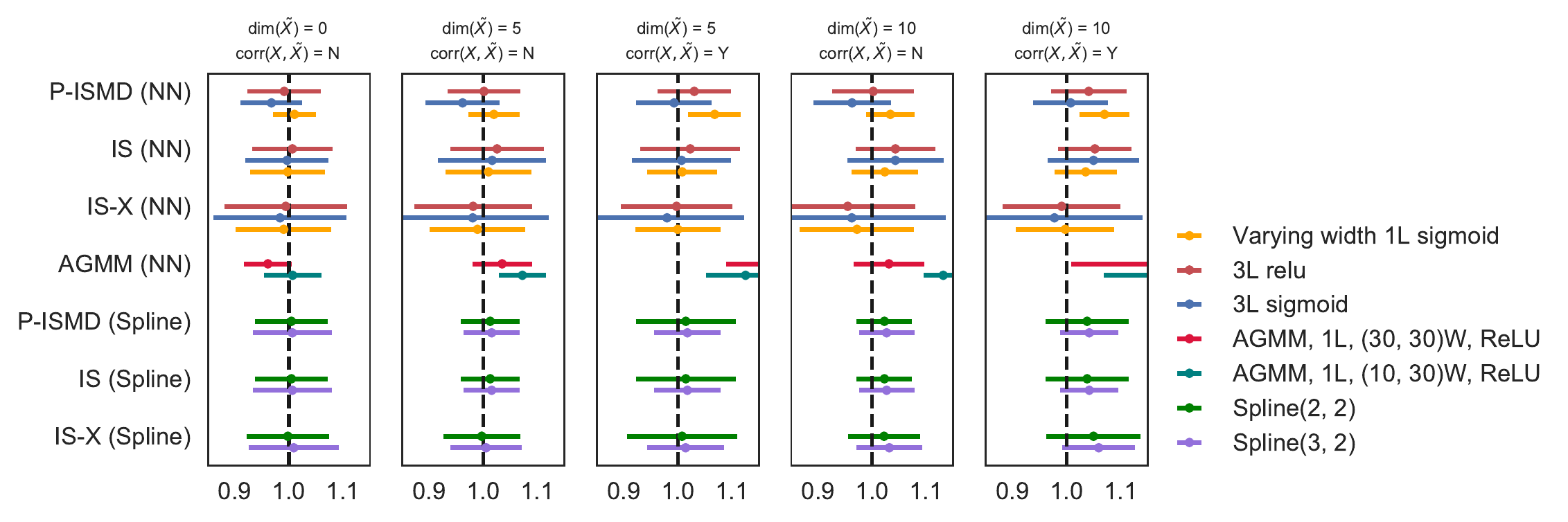}

    (b)  \cref{mc:2}, Nonparametric, $n=5000$

  \includegraphics[width=\textwidth]{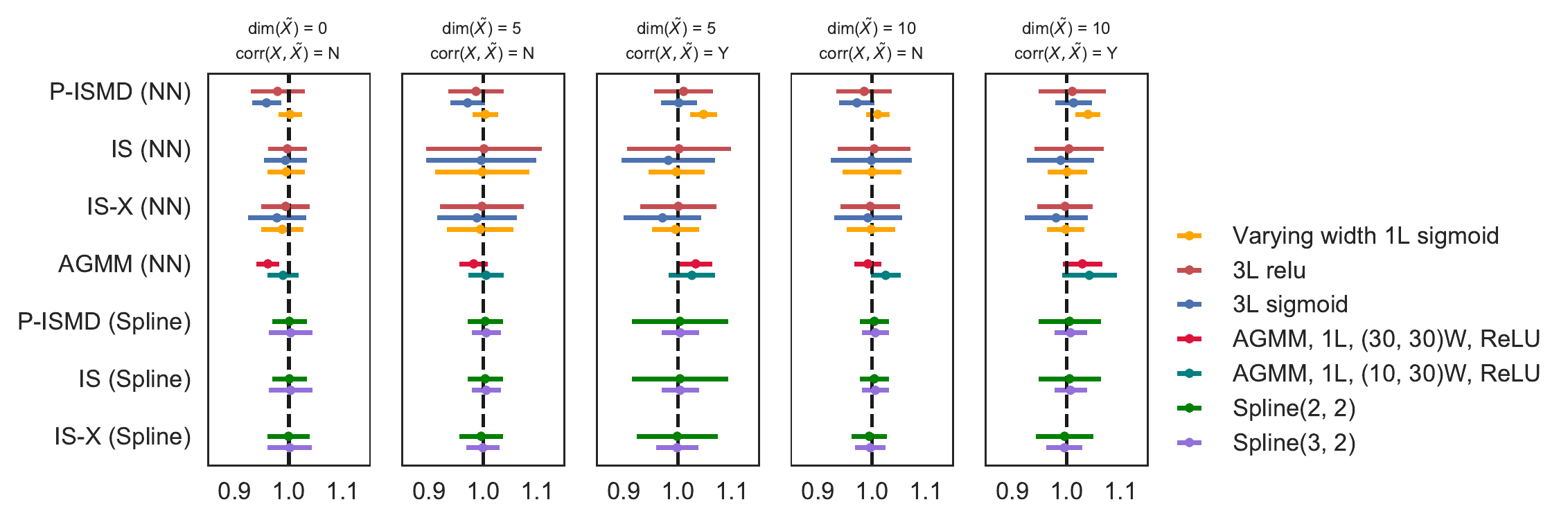}
  
       \begin{proof}[Notes]
       
       \footnotesize 
    
Monte Carlo Mean $\pm 1$ Monte Carlo standard deviation across
        1,000
replications. 

We consider a few different estimation strategies and vary over choice of
tuning parameters for nuisance parameters in these estimation strategies. 

In terms of estimation strategies, \is{} stands for identity
score estimators, detailed in
\cref{score_est}, whereas \isns-X stands for the score estimators,
but
with two-fold cross-fitting. AGMM uses the adversarial GMM
estimation algorithm in \cite{dikkala2020minimax} to compute $\hat h$, and
outputs the simple plug-in estimator for $\theta$. \ismd{}
estimators follow \cref{smd_est}. 

In terms of neural architecture and spline parameter choices, \emph{varying
width 1L sigmoid} refers
to using 1-layer sigmoid network, but vary the width of the network
according to $\dim(\tilde X)$, as opposed to fixing the width at 10. The
two AGMM architecture choices refer to different widths for the network
estimating $h$ and the adversarial network approximating the instrument
test functions, where (10,30)W refers to using width-10 for $h$ and
width-30 for the instruments. Lastly, $\text{Spline}(a,b)$ is a spline
basis for approximating $h$ such that each spline function is an $
(a-1)$-degree piecewise
polynomial that have $b$ knots, where we include pairwise interactions in
lieu of tensor products. In the spline scenarios, $\text{Spline}(a+1, b)$
is used as a spline basis for the instruments. Tuning parameter choices for
estimation of additional nuisance parameters are detailed in 
\cref{tab:nuisance}.
 \end{proof} 
  \label{fig:f3id}
\end{figure}

  \begin{figure}[htb]
  
  \caption{Estimation quality of average derivative parameter in 
  \cref{mc:2} across a variety of \emph{optimally weighted} estimators}
  \centering
  \ \ 
  
    (a)  \cref{mc:2}, Nonparametric, $n=1000$
    
  \includegraphics[width=\textwidth]{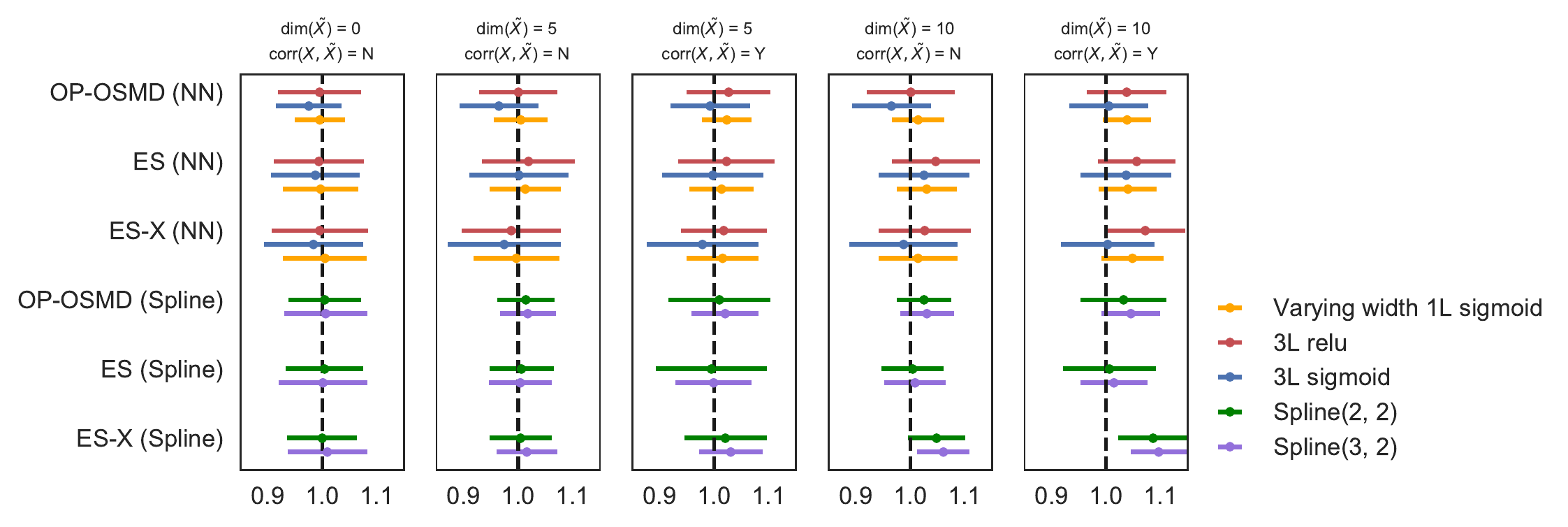}

    (b)  \cref{mc:2}, Nonparametric, $n=5000$

  \includegraphics[width=\textwidth]{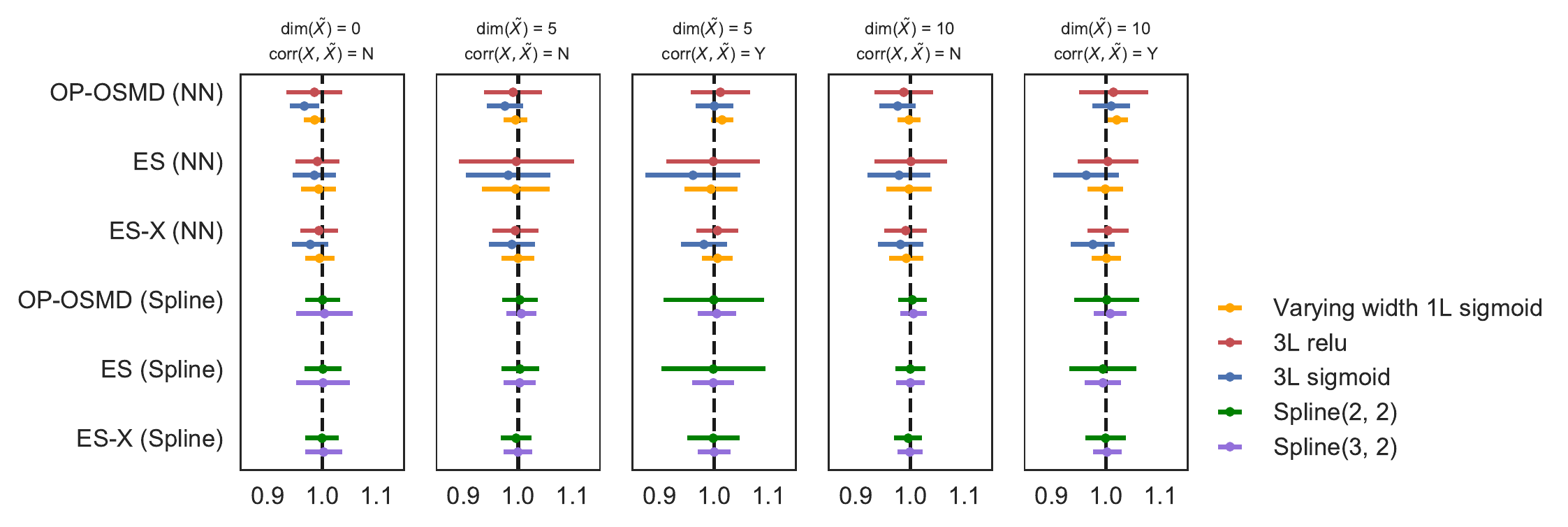}
  
       \begin{proof}[Notes]
       
       \footnotesize 
    
Monte Carlo Mean $\pm 1$ Monte Carlo standard deviation across
        1,000
replications. 

We consider a few different estimation strategies and vary over choice of
tuning parameters for nuisance parameters in these estimation strategies. 

In terms of estimation strategies, \es{} stands for efficient score estimators,
detailed in \cref{score_est}, whereas \esns-X stands for the score estimators,
but with two-fold cross-fitting. \osmd{} estimators follow \cref{smd_est}.

In terms of neural architecture and spline parameter choices, \emph{varying
width 1L sigmoid} refers to using 1-layer sigmoid network, but vary the width of
the network according to $\dim(\tilde X)$, as opposed to fixing the width at 10.
Lastly, $\text{Spline}(a,b)$ is a spline basis for approximating $h$ such that
each spline function is an $ (a-1)$-degree piecewise polynomial that have $b$
knots, where we include pairwise interactions in lieu of tensor products. In the
spline scenarios, $\text{Spline}(a+1, b)$ is used as a spline basis for the
instruments. Tuning parameter choices for estimation of additional nuisance
parameters are detailed in \cref{tab:nuisance}.
 \end{proof} 
  \label{fig:f3op}
\end{figure}

\begin{landscape}

    \begin{figure}[tb]
    \centering
    $ n = 1000$
    
    \includegraphics[width=1.3\textwidth]
    {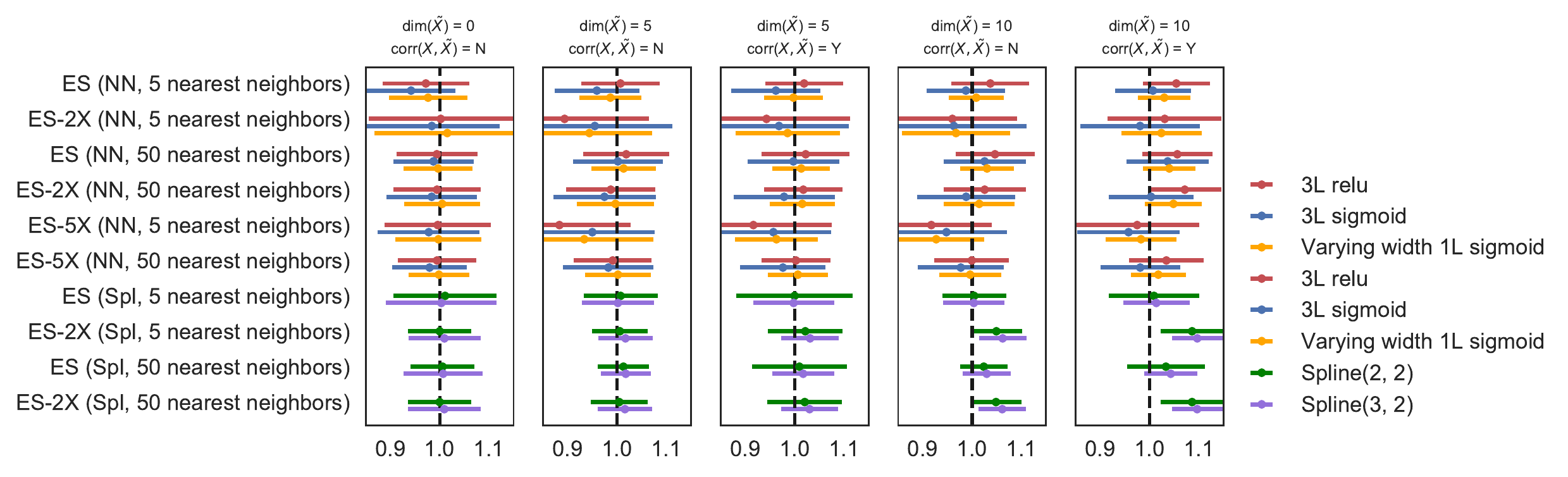}
    \caption{Performance of \es with different estimators for $\Sigma(X)^
    {-1}$ in the score expression}
    
    \label{fig:es_sensitivity}
      
  \begin{proof}[Notes]{
     Monte Carlo Mean $\pm 1$ Monte Carlo standard deviation across 1,000
replications. 

``$k$-nearest neighbors'': Use $k$-nearest neighbors to estimate $\Sigma
(X)$ in the score (in $\E[v^\star \mid X] \Sigma(X)^{-1}$).

``True inverse variance'': Plug in the true $\Sigma(X)$ for that in the
score.

``Plug in identity'': Plug in the identity matrix for $\Sigma(X)$ in the
score.

``Projection'': Use the projection of the squared residuals onto spline
bases for $\Sigma(X)$.

``Estimate $w^\star$'': Instead of estimating $v^\star$ with sieves, we estimate
$w^\star$ with sieves and form $v^\star$ via plugging in estimates of other
nuisance parameters. }
 \end{proof}
    \label{fig:es_sigma_inv}
\end{figure}

    \begin{figure}[tb]
    \centering
    $ n = 5000$
    
    \includegraphics[width=1.3\textwidth]
    {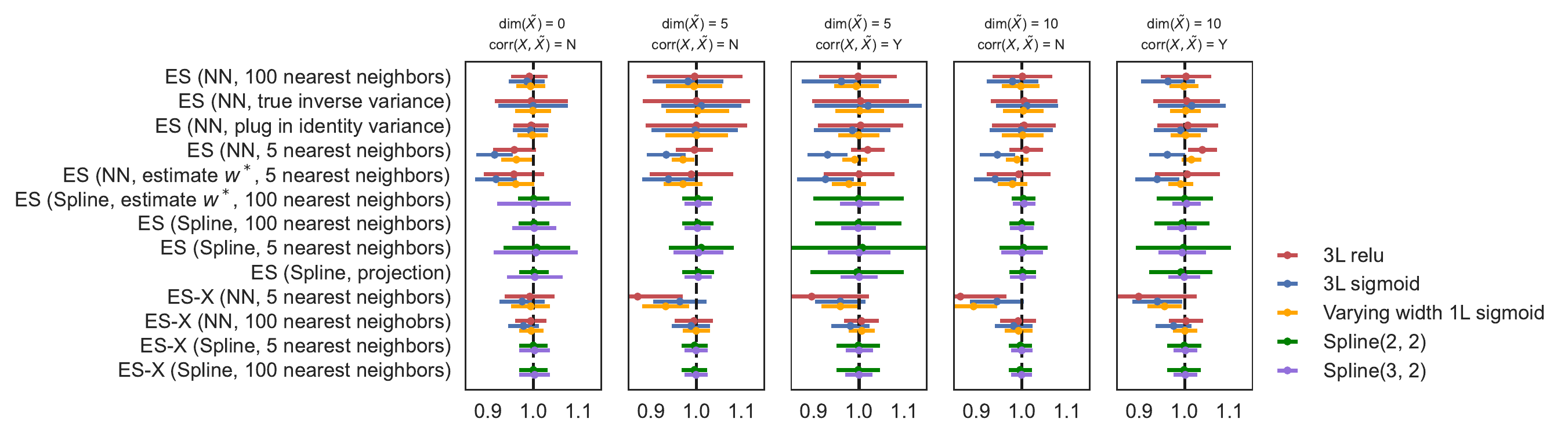}
    \caption{Performance of \es with different estimators for $\Sigma(X)^
    {-1}$ in the score expression}

  \begin{proof}[Notes]{
     Monte Carlo Mean $\pm 1$ Monte Carlo standard deviation across 1,000
replications. 

``$k$-nearest neighbors'': Use $k$-nearest neighbors to estimate $\Sigma
(X)$ in the score (in $\E[v^\star \mid X] \Sigma(X)^{-1}$).

``True inverse variance'': Plug in the true $\Sigma(X)$ for that in the
score.

``Plug in identity'': Plug in the identity matrix for $\Sigma(X)$ in the
score.

``Projection'': Use the projection of the squared residuals onto spline
bases for $\Sigma(X)$.

``Estimate $w^\star$'': Instead of estimating $v^\star$ with sieves, we estimate
$w^\star$ with sieves and form $v^\star$ via plugging in estimates of other
nuisance parameters. }
 \end{proof}
    \label{fig:es_sigma_inv5000}
\end{figure}

\begin{figure}
    \includegraphics[width=1.3\textwidth]{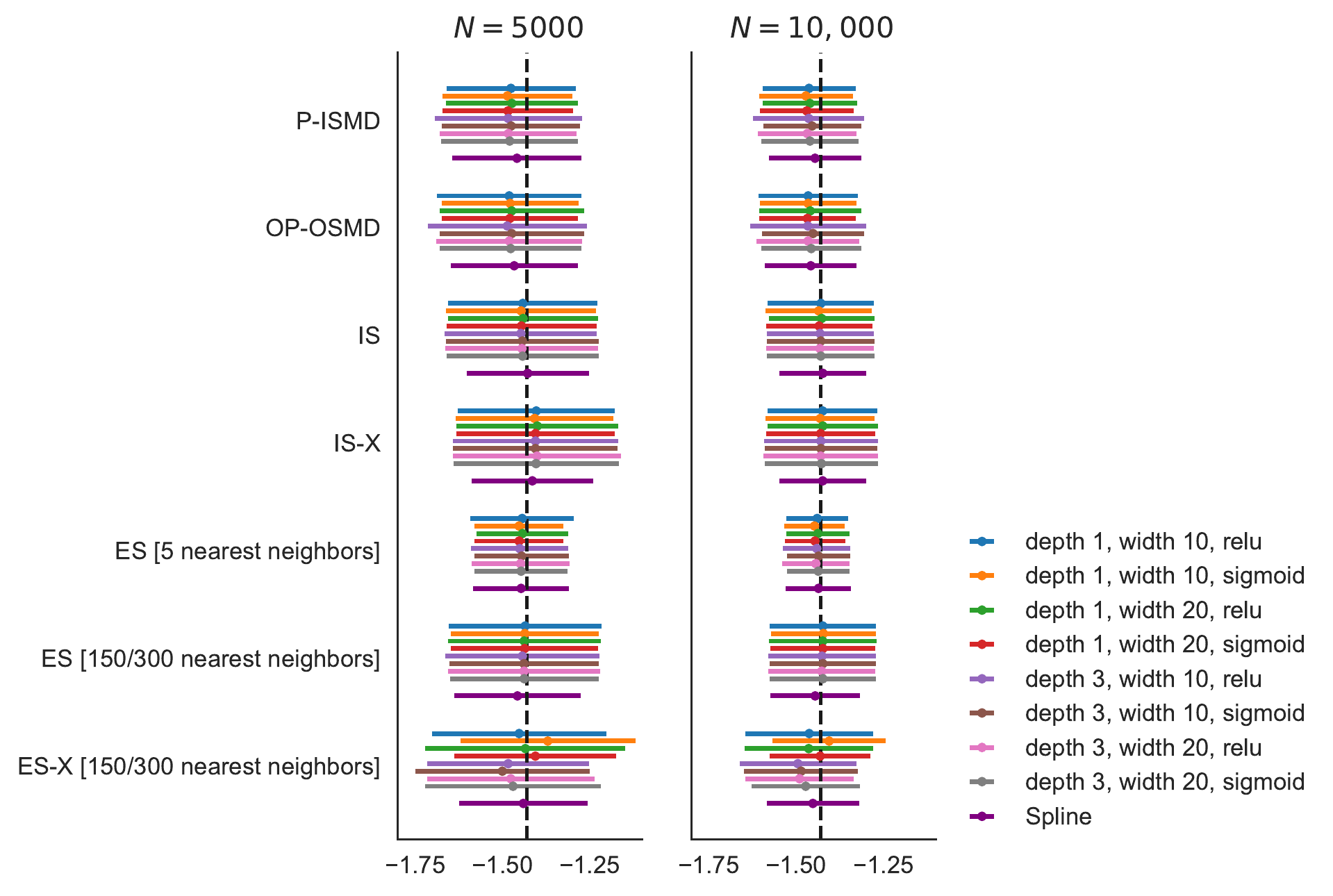}
    \caption{Performance of various estimators in \cref{mc:calibrated} across two
    sample sizes}
    \label{fig:calibrated}
    \begin{proof}[Notes]{
     Monte Carlo Mean $\pm 1$ Monte Carlo standard deviation across 1,000
replications. 

``150/300 nearest neighbors'': 150 nearest neighbors for $N=5000$, 300 nearest neighbors
for $N=10000$ in estimating $\hat \Sigma(X)$.

Splines use two knot cubic splines for instrument basis, and two knot quadratic splines
for endogenous functions.
 }
 \end{proof}
\end{figure}

\end{landscape}

\newpage

\begin{landscape}
    \begin{table}
    \footnotesize
    \begin{tabularx}{1.3\textwidth}{XXXXX}
    \toprule
    Estimator type & $\Sigma(X)$ [SMD] & $\Gamma(X)$ & $\Sigma(X)$ [Score]
    & $v^\star$ [Score] \\ \midrule
    P-ISMD [NN] & &&& \\ 
    OP-OSMD [NN] & 5 nearest neighbors & Projection of demeaned $(\nabla \hat h
    - \bar{\nabla \hat h})((y - \hat h) - \bar{(y - \hat h)})$ on
    instrument basis ($\phi$)
    used in SMD estimation. Then multiply $\Sigma(X)^{-1}$ estimate for
    SMD. Project the result onto the sieve basis for the instruments. & &
    \\
    \midrule 
    IS [NN] & & &  & Sieve calculation of \eqref{eq:w_star_id} with Spline(3,
    2) ($\lambda$)\\ 
    ES [NN] & Same as OP-OSMD  [NN] & Same as OP-OSMD [NN] & 50 nearest neighbors for
    $n=1000$, 100 for $n=5000$ & Sieve calculation of \eqref{eq:vstar_cl}
    with Spline(3, 2)  ($\lambda$)\\ \midrule 
    IS/ES-X [NN]& \multicolumn{4}{>
    {\hsize=\dimexpr3\hsize+4\tabcolsep+2\arrayrulewidth\relax}X}{Both scores take the form of $\nabla\hat h
    -
    \lambda
    (x)'\hat \xi \cdot (y - \hat h)$ where $\lambda(x)$ is a sieve basis 
    (Spline(3, 2)). The sample is split so that $\hat h$ and $\hat \xi$
    are estimated from one half and the score is computed on the other.
    The roles of the two subsamples are then exchanged.}\\
    \\\hline\hline\\
    P-ISMD [Spl] & &&& \\ 
    OP-OSMD [Spl] & Projection onto sieve basis $(\lambda)$ for the instruments &
    Projection
    of demeaned $(\nabla \hat h
    - \bar{\nabla \hat h})((y - \hat h) - \bar{(y - \hat h)})$ on
    instrument basis ($\lambda$)
    used in SMD estimation. Then multiply $\Sigma(X)^{-1}$ estimate from
    SMD.  Project the result onto the sieve basis for the instruments. & & \\
    \midrule 
    IS [Spl] & & &  & Sieve calculation of \eqref{eq:w_star_id} with same
    spline basis ($\lambda$) as the instruments\\
    ES [Spl] & Same as OP-OSMD [Spl] & Same as OP-OSMD [Spl] & 50 nearest neighbors for
    $n=1000$, 100 for $n=5000$ & Sieve calculation of \eqref{eq:vstar_cl}
    with same
    spline basis ($\lambda$) as the instruments \\ \midrule 
    IS/ES-X [Spl]& \multicolumn{4}{>
    {\hsize=\dimexpr3\hsize+4\tabcolsep+2\arrayrulewidth\relax}X}{Both
    scores take the form of $\nu(y_2)'\hat\beta
    -
    \lambda
    (x)'\hat \xi \cdot (y - \hat h)$ where $\lambda(x), \nu(y_2)$ are 
    sieve bases.
    The sample is split so that $\hat \beta$ and $\hat \xi$
    are estimated from one half and the score is computed on the other.
    The roles of the two subsamples are then exchanged.}\\
    \midrule     \end{tabularx}
    
    \caption{Estimation of additional nuisance parameters}
    \label{tab:nuisance}
\end{table}

\end{landscape}

\newpage 

\begin{landscape}
\thispagestyle{empty}
\begin{table}
\caption{SMD inference results for \ismd{} and \osmdns{} average derivative parameter in 
  \cref{mc:2}, $n=1000$} %
\label{tab:infe}

\ \ 

    \footnotesize
    \begin{tabular}{lllllcccccccccc}
\toprule
   &        &   &         & \multicolumn{5}{c}{P-ISMD} & \multicolumn{5}{c}{OP-OSMD} \\
   &        &   &         &   Mean &   Std & Med. Est. SE & Boot. LB & Boot. UB &    Mean &   Std & Med. Est. SE & Boot. LB & Boot. UB \\
Nui. Dim & Corr($X, \tilde X$) & Depth & Activation &        &       &              &          &          &         &       &              &          &          \\
\midrule
0  & 0.0000 & 1 & relu &  1.001 & 0.042 &        0.045 &    0.948 &    1.077 &   1.001 & 0.057 &        0.029 &    0.909 &    1.102 \\
   &        &   & sigmoid &  1.022 & 0.036 &        0.046 &    0.978 &    1.118 &   0.998 & 0.040 &        0.026 &    0.913 &    1.108 \\
   &        & 3 & relu &  0.991 & 0.068 &        0.053 &    0.913 &    1.087 &   0.995 & 0.076 &        0.033 &    0.840 &    1.235 \\
   &        &   & sigmoid &  0.968 & 0.057 &        0.069 &    0.871 &    1.107 &   0.975 & 0.060 &        0.043 &    0.883 &    1.126 \\
5  & 0.0000 & 1 & relu &  1.016 & 0.048 &        0.054 &    0.889 &    1.200 &   1.009 & 0.061 &        0.040 &    0.870 &    1.218 \\
   &        &   & sigmoid &  1.022 & 0.048 &        0.057 &    0.873 &    1.103 &   1.005 & 0.050 &        0.037 &    0.893 &    1.133 \\
   &        & 3 & relu &  1.001 & 0.068 &        0.061 &    0.878 &    1.085 &   1.000 & 0.072 &        0.043 &    0.875 &    1.096 \\
   &        &   & sigmoid &  0.962 & 0.069 &        0.088 &    0.713 &    1.024 &   0.964 & 0.072 &        0.061 &    0.722 &    1.048 \\
   & 0.5000 & 1 & relu &  1.052 & 0.049 &        0.052 &    0.946 &    1.159 &   1.033 & 0.055 &        0.038 &    0.906 &    1.142 \\
   &        &   & sigmoid &  1.067 & 0.048 &        0.053 &    0.921 &    1.163 &   1.023 & 0.046 &        0.035 &    0.904 &    1.116 \\
   &        & 3 & relu &  1.030 & 0.068 &        0.060 &    0.895 &    1.143 &   1.026 & 0.077 &        0.043 &    0.905 &    1.167 \\
   &        &   & sigmoid &  0.993 & 0.070 &        0.090 &    0.740 &    1.054 &   0.992 & 0.073 &        0.062 &    0.722 &    1.072 \\
10 & 0.0000 & 1 & relu &  1.027 & 0.052 &        0.061 &    0.915 &    1.090 &   1.013 & 0.051 &        0.045 &    0.908 &    1.104 \\
   &        &   & sigmoid &  1.027 & 0.046 &        0.062 &    0.900 &    1.111 &   1.012 & 0.048 &        0.042 &    0.940 &    1.157 \\
   &        & 3 & relu &  1.002 & 0.076 &        0.070 &    0.749 &    1.179 &   1.001 & 0.081 &        0.052 &    0.753 &    1.184 \\
   &        &   & sigmoid &  0.963 & 0.072 &        0.101 &    0.670 &    1.015 &   0.965 & 0.073 &        0.072 &    0.682 &    1.026 \\
   & 0.5000 & 1 & relu &  1.064 & 0.078 &        0.064 &    0.968 &    1.147 &   1.041 & 0.057 &        0.045 &    0.959 &    1.148 \\
   &        &   & sigmoid &  1.071 & 0.047 &        0.064 &    0.931 &    1.133 &   1.037 & 0.046 &        0.044 &    0.935 &    1.135 \\
   &        & 3 & relu &  1.041 & 0.070 &        0.073 &    0.865 &    1.251 &   1.038 & 0.073 &        0.052 &    0.867 &    1.278 \\
   &        &   & sigmoid &  1.007 & 0.070 &        0.105 &    0.749 &    1.041 &   1.005 & 0.073 &        0.075 &    0.740 &    1.056 \\
\bottomrule
\end{tabular}

    \begin{proof}[Notes]
        1000 Monte Carlo replications. Bootstrap CIs based on a single
        replication. 
    \end{proof}

\end{table}

\begin{table}
\caption{SMD inference results for \ismd{} and \osmdns{} average derivative parameter in 
  \cref{mc:2}, $n=5000$} %
\label{tab:infe2}

\ \ 

    \footnotesize
    \begin{tabular}{lllllcccccccccc}
\toprule
   &        &   &         & \multicolumn{5}{c}{P-ISMD} & \multicolumn{5}{c}{OP-OSMD} \\
   &        &   &         &   Mean &   Std & Med. Est. SE & Boot. LB & Boot. UB &    Mean &   Std & Med. Est. SE & Boot. LB & Boot. UB \\
Nui. Dim & Corr($X, \tilde X$) & Depth & Activation &        &       &              &          &          &         &       &              &          &          \\
\midrule
0  & 0.0000 & 1 & relu &  0.985 & 0.022 &        0.021 &    0.966 &    1.036 &   0.994 & 0.023 &        0.013 &    0.897 &    1.031 \\
   &        &   & sigmoid &  1.013 & 0.018 &        0.021 &    0.976 &    1.054 &   0.990 & 0.018 &        0.011 &    0.942 &    1.026 \\
   &        & 3 & relu &  0.979 & 0.050 &        0.023 &    0.933 &    1.012 &   0.986 & 0.051 &        0.014 &    0.850 &    1.056 \\
   &        &   & sigmoid &  0.958 & 0.027 &        0.033 &    0.876 &    0.978 &   0.967 & 0.027 &        0.021 &    0.881 &    1.035 \\
5  & 0.0000 & 1 & relu &  0.996 & 0.022 &        0.024 &    0.968 &    1.079 &   0.997 & 0.025 &        0.017 &    0.912 &    1.076 \\
   &        &   & sigmoid &  1.006 & 0.024 &        0.025 &    0.964 &    1.065 &   0.995 & 0.022 &        0.015 &    0.940 &    1.034 \\
   &        & 3 & relu &  0.986 & 0.051 &        0.026 &    0.922 &    1.041 &   0.990 & 0.053 &        0.018 &    0.914 &    1.043 \\
   &        &   & sigmoid &  0.971 & 0.032 &        0.037 &    0.938 &    1.071 &   0.976 & 0.033 &        0.024 &    0.933 &    1.078 \\
   & 0.5000 & 1 & relu &  1.028 & 0.025 &        0.023 &    0.976 &    1.071 &   1.014 & 0.025 &        0.016 &    0.907 &    1.052 \\
   &        &   & sigmoid &  1.046 & 0.024 &        0.022 &    1.002 &    1.117 &   1.014 & 0.021 &        0.014 &    0.960 &    1.049 \\
   &        & 3 & relu &  1.011 & 0.054 &        0.025 &    0.942 &    1.045 &   1.011 & 0.055 &        0.018 &    0.923 &    1.038 \\
   &        &   & sigmoid &  1.002 & 0.033 &        0.037 &    0.927 &    1.066 &   1.000 & 0.034 &        0.025 &    0.908 &    1.060 \\
10 & 0.0000 & 1 & relu &  0.999 & 0.033 &        0.025 &    0.959 &    1.044 &   0.994 & 0.025 &        0.018 &    0.933 &    1.031 \\
   &        &   & sigmoid &  1.005 & 0.022 &        0.025 &    0.969 &    1.064 &   0.997 & 0.022 &        0.016 &    0.936 &    1.032 \\
   &        & 3 & relu &  0.985 & 0.052 &        0.027 &    0.848 &    1.050 &   0.988 & 0.054 &        0.019 &    0.848 &    1.055 \\
   &        &   & sigmoid &  0.972 & 0.033 &        0.039 &    0.930 &    1.064 &   0.977 & 0.034 &        0.027 &    0.932 &    1.068 \\
   & 0.5000 & 1 & relu &  1.029 & 0.056 &        0.025 &    0.981 &    1.065 &   1.016 & 0.023 &        0.018 &    0.956 &    1.047 \\
   &        &   & sigmoid &  1.042 & 0.025 &        0.024 &    1.008 &    1.113 &   1.020 & 0.021 &        0.016 &    0.973 &    1.065 \\
   &        & 3 & relu &  1.010 & 0.062 &        0.028 &    0.911 &    1.126 &   1.013 & 0.063 &        0.020 &    0.901 &    1.122 \\
   &        &   & sigmoid &  1.013 & 0.034 &        0.041 &    0.937 &    1.073 &   1.009 & 0.035 &        0.028 &    0.927 &    1.065 \\
\bottomrule
\end{tabular}

    \begin{proof}[Notes]
        1000 Monte Carlo replications. Bootstrap CIs based on a single
        replication. 
    \end{proof}

\end{table}

\end{landscape}

\begin{figure}[htb]

  \caption{Inference quality of average derivatrive parameter in 
  \cref{mc:2} across a variety of estimators}
  \label{fig:f4}
  \ \ 
  
  \centering
   (a)  \cref{mc:2}, Nonparametric, $n=1000$
    
  \includegraphics[width=\textwidth]{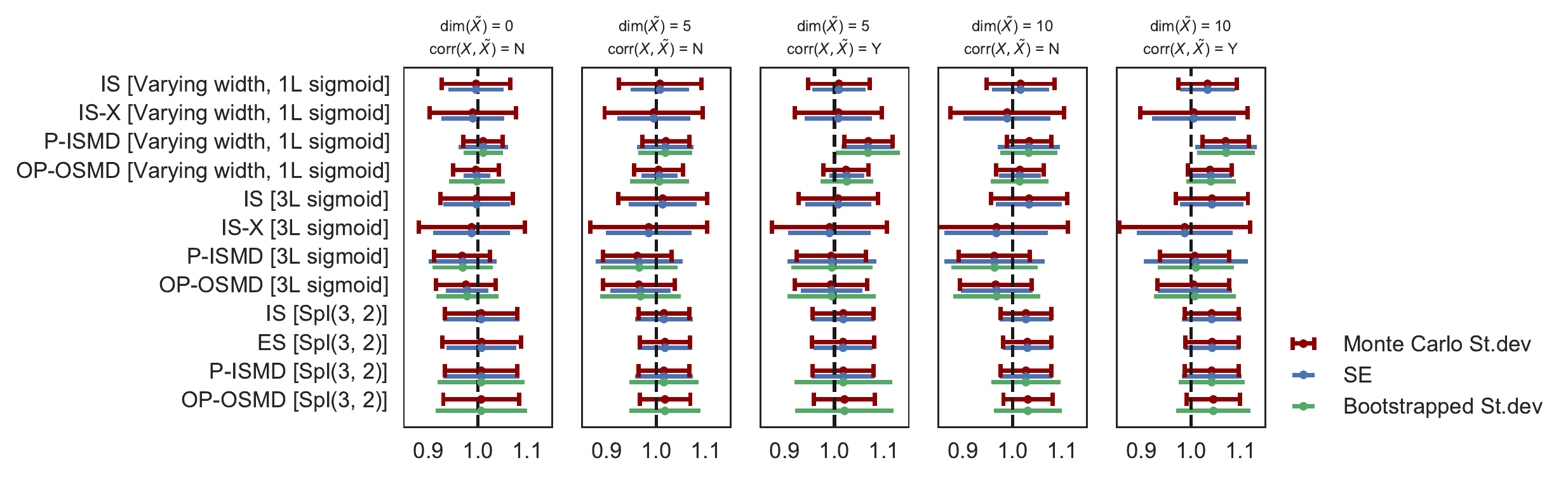}
    
    \ \ \ 
    
  (b)  \cref{mc:2}, Nonparametric, $n=5000$
    
  \includegraphics[width=\textwidth]{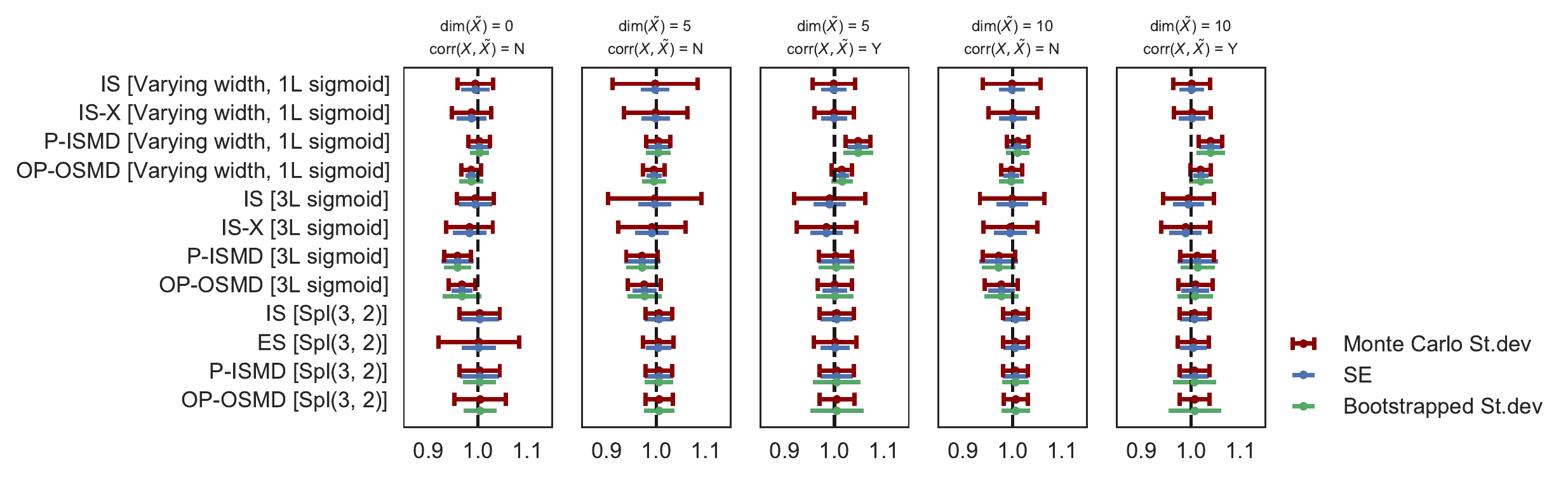}
  
         \begin{proof}[Notes]{\footnotesize
     Monte Carlo Mean $\pm 1 \{\text{Monte Carlo st. dev.,
     estimated s.e., bootstrapped s.e.}\}$ across
     1,000
replications. 

Bootstrap SEs are based on one realization of the data. }
\end{proof} 
\end{figure}

\appendix

\section{Appendix: Additional Monte Carlo Results}
\label{sub:amc}

\noindent {\bf An additional Monte Carlo and sensitivity to instrument basis.} 

\begin{mc}
\label{mc:1}
     This is an augmentation of the design
in \cite{chen2007large}. This design has a simpler functional form of $h_0$:
\[
Y_1 =h_0(Y_2)+U= X_1  + h_{01}(R) + h_{02}(X_2) + h_{03}(\tilde X) + U, \quad
\E[U
\mid
X_1,X_2,X_3, \tilde X] = 0,
\]
where we generate
\begin{align*}
h_{01} : \R \to \R &\quad t \mapsto\frac{1}{1+\exp(-t)}\\ 
h_{02}(t) : \R \to \R & \quad t \mapsto \log (1+t) \\
h_{03}: \R^{d_{\tilde x}} \to \R &\quad \tilde x \mapsto 5\tilde x_1^3 +
\tilde x_2 \cdot \max_{j=1,\ldots, d_{\tilde x}} \pr{\tilde
x_j\maxwith 0.5} + 0.5\exp(-\tilde x_{d_{\tilde x}}) \\
X_1, X_2, X_3
&\sim \Unif[0,1] \\
U \mid X_1,X_2,X_3 &\sim \Norm\pr{0,\frac{1}{3}
(X_1^2+X_2^2 + X_3^2)}\\
\epsilon &\sim \Norm(0,0.1) \\ R &= X_1
+ X_2 + X_3 + 0.9U + \epsilon.
\end{align*}
The process generating $\tilde X$ is
somewhat complex. First, we generate a covariance matrix $\Sigma \propto (I
+ Z'Z)$, normalized to unit diagonals, where $Z$'s entries are i.i.d.
standard Normal. The seed generating the covariance matrix is held fixed
over different samples, and so $\Sigma$ should be viewed as fixed a
priori. Next, let $\rho \in [-1,1]$ denote a
correlation level and we let \begin{equation}
\tilde X = \Phi\pr{\rho (X_1 + X_2 + X_3)  + \sqrt{1-\rho^2} T}\quad T \sim
\Norm(0,\Sigma),
    \label{eq:tildex}
\end{equation}
where $\Phi(\cdot)$ is the standard Normal CDF, and $\Phi(\cdot)$ and
addition are applied elementwise. In the exercises reported, we use $\rho
\in \{0, 0.5\}$ for correlation levels. This Monte Carlo design becomes
identical to the one used in 
\cite{chen2007large}  when $\tilde X$ is an empty vector. We
increase the dimension of $\tilde X$ to $5$ and $10$ to make the estimation problem more
difficult. Note that this design allows for correlation among regressors both endogenous and exogenous. It also allows for heteroskedasticity and possibly large dimensions by increasing the dimension of $\tilde X$. We have also tried different 
conditional variance of $U$, the simulation results are similar.

 \noindent To connect with the notation in the previous sections, let $Y_{2} = [X_1, R, X_2, \tilde X]$ and $X = [X_1, X_2, X_3,
 \tilde X]$. The parameter of interest is $\theta_0 = \E\bk{\diff{h_0(Y_2)}{X_1}}=1$.
\end{mc}

\begin{figure}[htb]
  \centering  
  \caption{ANN SMD estimators for the average derivative parameter in 
  \cref{mc:1}}
\includegraphics[width=\textwidth]
{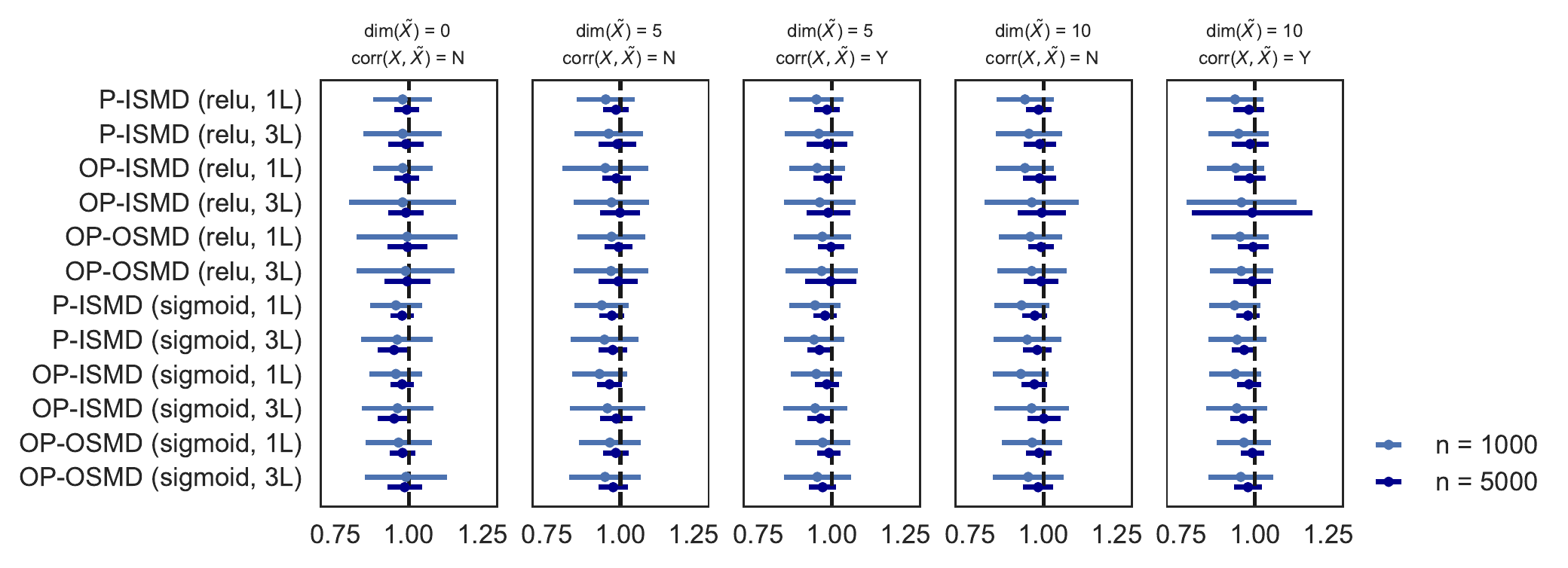}

\begin{proof}[Notes]{\footnotesize
     Monte Carlo Mean $\pm 1$ Monte Carlo standard deviation across 1,000
replications.}
 \end{proof} 
  
  \label{fig:f1}
\end{figure}

The estimation choices for \cref{mc:1} are exactly as in \cref{smd_est}. \Cref{fig:f1}
plots the performance of various ANN SMD estimators in terms of mean $\pm$ one (Monte
Carlo) standard deviation across 1000 replications for \cref{mc:1}, in which the first
element of $Y_2$ is exogenous ($X_1$). As a reminder, \ismd is the simple plug in
estimator of $\theta$ with identity weighting, while \osmd is the orthogonalized plug in
with optimal weighting for the SMD objective. As we can see across layers and activation
function, and whether we have a low dimensional regime in the left hand side columns or
large dimensional regimes in the right hand columns, or whether there is correlation
across regressors (denoted by Y(es) or N(o) on top of each column), the behavior of these
ANN estimators is similar and adequate. All the intervals are more or less centered on top
of the truth, $\theta_0=1$, while the efficient estimator \osmd is slightly less biased.

\begin{figure}[tb]
 
  \caption{ANN SMD estimators for \cref{mc:1} with smaller instrument
  basis}
  \label{fig:f1-smaller-instrument}
  \centering

  \includegraphics[width=\textwidth]{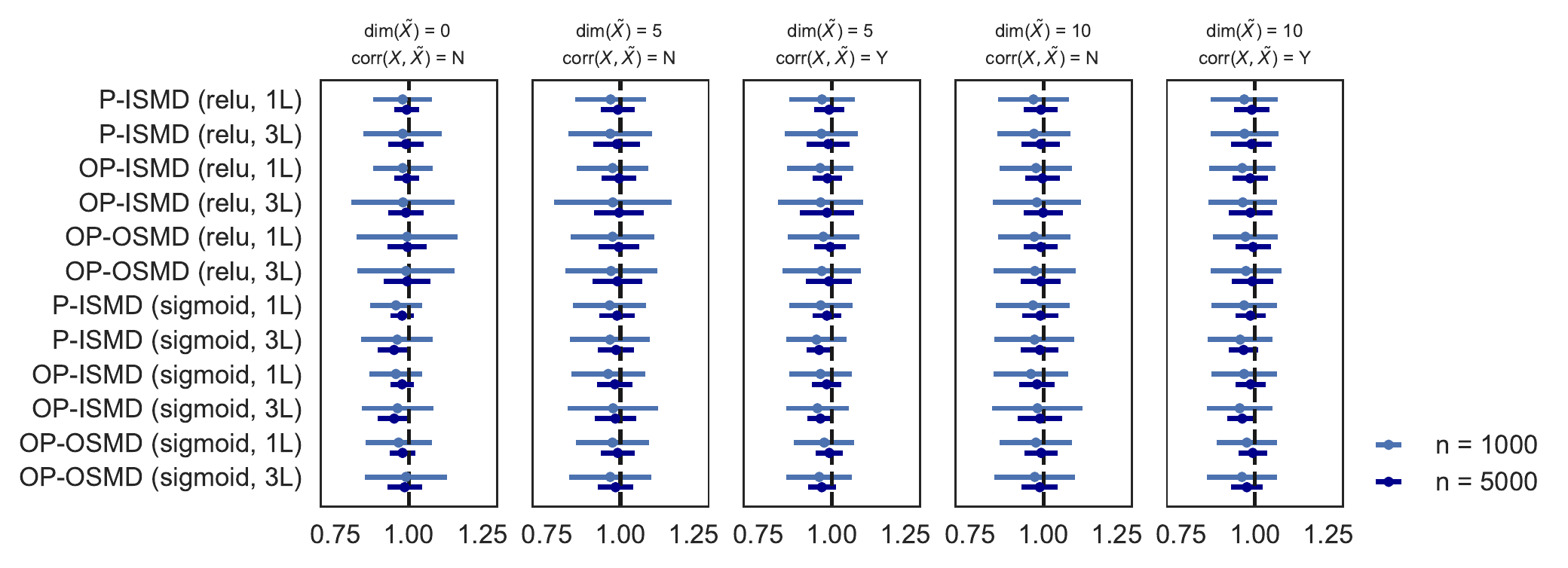}

  \begin{proof}[Notes]{\footnotesize
     Monte Carlo Mean $\pm 1$ Monte Carlo standard deviation across 1,000
replications. 

}
 \end{proof}
\end{figure}

\begin{figure}[tb]
 
  \caption{ANN SMD estimators for \cref{mc:2} with smaller instrument
  basis}
  \label{fig:f2-smaller-instrument}
  \centering
  
  \ \ 
  
  (a) $n = 1000$ (Smaller instrument basis)
  
  \includegraphics[width=\textwidth]{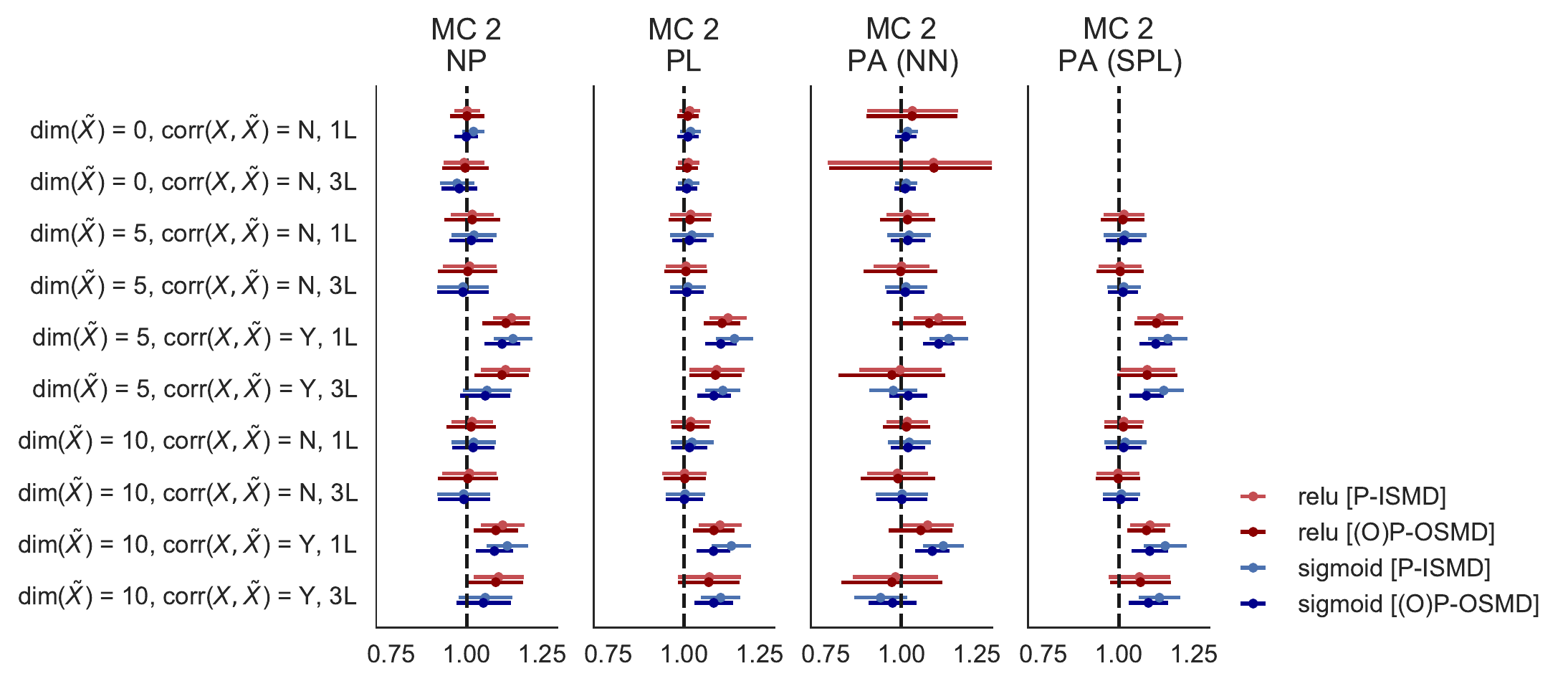}

  \ \ 
  
  (b) $n = 5000$ (Smaller instrument basis)
  
  \includegraphics[width=\textwidth]{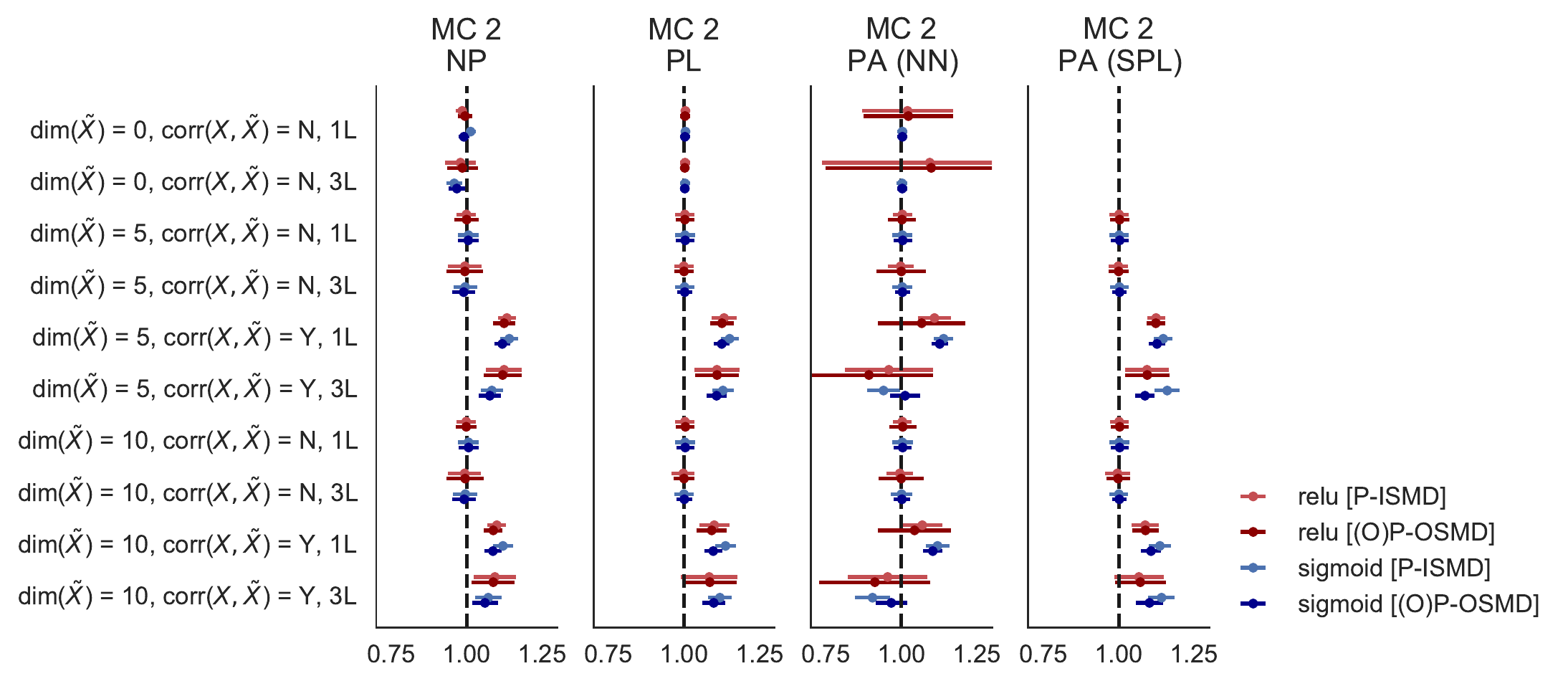}
  
  \begin{proof}[Notes]{
\footnotesize
     Monte Carlo Mean $\pm 1$ Monte Carlo standard deviation across 1,000
replications. 

The columns are estimators where different correct assumptions of
the data-generating process are placed. The first column (NP: 
\emph{nonparametric}) shows estimated average derivative of an NPIV model,
where the unknown function $h(Y_2)$ is not assumed to have separable
structure. The second
column (PL: \emph{partially linear}) assumes $h(Y_2) = \theta R_1 + h_1
(R_2, X_2, \tilde X)$. The third and fourth columns (PA: \emph{partially
additive}) assumes $h(Y_2)= \theta R_1 + h_1(R_2) + h_2(X_2) + h_3(\tilde
X)$. The third column uses neural networks to approximate the scalar
functions $h_1, h_2$, and the fourth column uses splines to approximate
$h_1, h_2$.

For each type of assumption placed on the true $h_0(Y_2)$, we vary the
data-generating process by varying the dimension of $\tilde X$ and the
level of
correlation between $(X_1,X_2,X_3)$ and $\tilde X$. We also vary the
network architecture by $\{\text{ReLU}, \text{Sigmoid}\} \times \{
\text{1L, 3L}\}.$ Lastly, we vary the type of estimator used from simple
plug-in with the identity-weighted SMD estimator to orthogonalized plug-in
with the optimally-weighted SMD estimator.}
 \end{proof}
\end{figure}

\cref{fig:f1-smaller-instrument,fig:f2-smaller-instrument} are replicates of
 \cref{fig:f1,fig:f2} respectively, except for slightly smaller instrument sieve bases.
 Specifically, we only use $\phi_1$ in item 1(a)i in \cref{sub:planned} as the basis, as
 opposed to using both $\phi_1,\phi_2$. We see that the ANN SMD estimates for the simpler
 \cref{mc:1} are not sensitive to the choice of instrument sieves, while the ANN SMD
 estimates for \cref{mc:2} are slightly more sensitive to the choice of instrument sieves.

\ \ \

\section{Appendix: Analysis of the optimally weighted SMD as sequential GMM}
\label{sub:cl}

Recall that we can view the SMD estimator as a plug-in: \[
\hat\theta -\theta_0 = \frac{1}{n}\sum_{i=1}^n \left[a(Y_{2i})\nabla_1 \hat h(Y_{2i}) -
\theta_0 -\hat{\Gamma}(X_i)
(Y_{1i} -
\hat h(Y_{2i}))\right]
\]
Again, we linearize \[
\hat \theta - \theta_0 \approx \frac{1}{n}\sum_{i=1}^n \left[a(Y_{2i})\nabla_1 h_0(Y_{2i}) - \theta_0
- \Gamma
(X_i)(Y_{1i} - h_0(Y_{2i})) + \frac{d \E[\varepsilon(Z, \alpha_0)]}{dh}[\hat h
- h_0]\right]
\]
and define \[
v \mapsto \frac{d \E[\varepsilon(Z, \alpha_0)]}{dh}[v]
\]
as a linear operator which admits a Riesz representation under the inner
product for the first-step SMD estimation:\[
\ip{u, v}_{\rho_2} = \E\bk{
   \frac{d\E[Y_1 - h_0(Y_2) \mid X]}{dh}[u]' \Sigma(X)^{-1} \frac{d\E[Y_1 - h_0
   (Y_2) \mid X]}{dh}[v]
}  = \E\bk{
    \E[u \mid X] \Sigma(X)^{-1} \E[v \mid X]
},
\]
since the pathwise derivative is\[
\frac{d\E[Y_1 - h_0(Y_2) \mid X]}{dh}[v] = -\E[v \mid X]. 
\]
Note that $\Sigma(X)$ is the scalar variance $\var(Y_1 - h_0(Y_2)|X)$.
Let $v^\star_{\rho_2}$
be the Riesz representer. Applying \cite{CL2015gmm} we obtain the asymptotic influence function expansion:
\begin{equation}\label{eq:ES}
\hat\theta -\theta_0  \approx \frac{1}{n}\sum_{i=1}^n \left[a(Y_{2i})\nabla_1 h_0(Y_{2i}) -
\theta_0 - \Gamma(X_i)
(Y_{1i} -
 h_0(Y_{2i})) + \frac{\E[v^\star_{\rho_2} \mid X]}{\Sigma(X)} (Y_1 - h_0
(Y_2))\right],
\end{equation}
which also verifies that $v_{\rho_2}^\star = v_h^\star$ are the same
object.\footnote{Assuming \emph{completeness}: $\E[h(Y_2) \mid X] = 0
\iff h(Y_2) = 0$ a.s.}

This alternative analysis allows us to estimate $v_
{\rho_2}^\star$ directly, instead of estimating $w^\star$ as
in the optimal weighed SMD case, since it shows that $v_{\rho_2}^\star = v_h^\star$ is
in fact a Riesz representer on its own with respect to a different inner
product. The definition of the
Riesz representer is such
that \[
\E[\E[v_{\rho_2}^\star \mid X] \Sigma(X)^{-1} \E[v \mid X]] = \E[a(Y_2)\nabla_1
v + \Gamma(X) v] \quad \norm{v^\star_{\rho_2}}_{\rho_2}^2 = \sup_v \frac{(\E[a(Y_2)\nabla_1
v + \Gamma(X) v])^2}{\E[ \Sigma(X)^{-1} \E[v \mid X]^2 ]}.
\]
Though this population version is difficult to characterize, it again can be approximated via linear sieve $\mathcal{V}_n =\{ \nu(Y_2)'\gamma: \gamma \}$ (for instance one can think of $\nu(Y_2)$ as a power series or splines or Fourier series in $Y_2$)
\[
\norm{v^\star_{\rho_2,n}}_{\rho_2}^2 = \sup_{v\in \mathcal{V}_n} \frac{(\E[a(Y_2)\nabla_1
v + \Gamma(X) v])^2}{\E[ \Sigma(X)^{-1} \E[v \mid X]^2 ]}
\]
then the sieve version of the Riesz representer is easy to
compute.
For completeness, the sieve Riesz representer is\[
v^\star_{\rho_2, n} = \nu(Y_2)' \E[\Sigma(X)^{-1} \E[\nu \mid X] \E[\nu \mid
X]']^{-1} \E[a(Y_2)\nabla_1 \nu + \Gamma(X) \nu]
\]
as a specialization of \cite{CL2015gmm}.

The root-$n$ asymptotic normality now can be established by checking the sufficient conditions in \cite{CL2015gmm}.

\end{document}